\documentclass[showpacs,amsmath,amssymb,aps,showkeys,floatfix,prc,onecolumn,a4paper,preprint]{revtex4}
\usepackage{graphicx}
\usepackage{dcolumn}
\usepackage{bm}
\usepackage{epsfig}
\usepackage{amsfonts}
\usepackage{amssymb,amscd}
\usepackage{hyperref}
\usepackage{xcolor}
\hypersetup{
    colorlinks=true,                
    breaklinks=true,                
    urlcolor= black,                
    linkcolor= blue,                
    bookmarksopen=false,
    filecolor=black,
    citecolor=blue,
    linkbordercolor=blue,
    pdfborder = {0 1 0}
}

\begin{document}

\title{$D$-meson production in high energy $pA$ collisions within the QCD color dipole transverse momentum representation}

\pacs{12.38.-t; 13.60.Le; 13.60.Hb}

\author{G. Sampaio dos Santos}
\affiliation{High Energy Physics Phenomenology Group, GFPAE  IF-UFRGS \\
Caixa Postal 15051, CEP 91501-970, Porto Alegre, RS, Brazil}

\author{G. Gil da Silveira}
\affiliation{CERN, PH Department, 1211 Geneva, Switzerland}
\affiliation{High Energy Physics Phenomenology Group, GFPAE  IF-UFRGS \\
Caixa Postal 15051, CEP 91501-970, Porto Alegre, RS, Brazil}

\author{M. V. T. Machado}
\affiliation{High Energy Physics Phenomenology Group, GFPAE  IF-UFRGS \\
Caixa Postal 15051, CEP 91501-970, Porto Alegre, RS, Brazil}

\begin{abstract}
The $D$-meson production is investigated by considering the unintegrated gluon distribution within the dipole approach in the momentum representation. We analyze the $D$-meson spectrum accounting  for the effects of nonlinear behavior of the QCD dynamics which can be accordingly addressed in the dipole framework. The unintegrated gluon distribution is obtained by using geometric scaling property and the results are compared to the Glauber-Gribov framework. The absolute transverse momentum spectra and the nuclear modification ratios are investigated.  Predictions are compared with the experimental measurements by the ALICE and LHCb Collaborations in $pA$ collisions for different rapidity bins. 
\end{abstract}

\maketitle

\section{Introduction}
\label{intro}

The heavy flavor production has been  extensively studied along the last years and robust theoretical formalisms have been  developed, especially with the establishment of the high energy particle accelerators, such as the Large Hadron Collider (CERN-LHC). As a consequence, the precision of the measurements in conjunction with a wide window of center-of-mass energy, transverse momentum, and rapidity distributions offer interesting prospects to investigate heavy quarks and heavy meson productions. In particular, the $D$-meson production may be considered as an useful source for investigating the heavy quarks and their interactions \cite{godwiss}. The heavy quark mass is large enough to be taken as a hard scale, allowing to evaluate the production cross sections via perturbative methods, such as the framework of perturbative Quantum Chromodynamics (pQCD) \cite{nason,mangano,cacciari,frixione}. The $D$-mesons in the final state are produced by the hadronization process of these heavy quarks. Hence, the charmed meson production may carry information about the heavy flavor fragmentation function \cite{anderle} as well as the partonic distribution in the nucleon or nucleus \cite{gauld,gauld1,kusina,eskola}. The relatively low mass of D-mesons allows investigations based on parton saturation effects mostly in the low-$p_T$ region. At forward rapidities the nuclear saturation scale, $Q_{s,A}$, is sufficiently high and should control the suppression in the nuclear modification factor, $R_{pA}^D$. 

From theoretical point of view,  the $D$-meson production cross section is described within the collinear factorization \cite{collins} or the $k_T$-factorization approach \cite{gribov,march,levin,catani}. Examples of pQCD calculations are  the general-mass variable-flavour-number scheme (GM-VFNS) \cite{kniehl,helenius} and the fixed order plus next-to-leading logarithms approach (FONLL) \cite{cacciari,cacciari1,cacciari2}. Investigations in the context of  $k_T$-factorization framework are found in Refs.~\cite{ryskin,kharz,shab,lusz,mac,chachamis,vicnav,carv}.
On the experimental side, the ALICE Collaboration at the CERN-LHC has recently reported the measurements of the azimuthal-correlation
function of prompt $D$-mesons with charged particles \cite{ALICE,ALICE1} and the measurements of prompt $D$-meson production as a function of multiplicity \cite{ALICE2} in $pp/pA$ collisions at $\sqrt{s} = 5.02$~TeV. The measurement of angular correlations is a powerful tool to investigate collective effects in a complex system as heavy-ion collisions. The measurements provided by the ALICE experiment are obtained by non-central collision events where an anisotropy is introduced in the angular distribution. In particular, saturation effects may modified the azimuthal correlation and be a window to investigate the saturation physics.

Charmed meson hadroproduction in the $k_T$-factorization approach is calculated by considering the unintegrated gluon distribution (UGD) which includes the transverse momenta of the initial partons. The UGDs can be parametrized, where their rapidity dependence, $Y=\ln (1/x)$, and gluon transverse momentum, $k_{\perp}$, vary depending on the underlying physical assumptions.  Consequently, observables strongly dependent on gluon initiated processes are crucial to constrain the UGDs in nucleons \cite{angeles}. On the other hand, cold nuclear matter effects are present in collisions involving a nucleus. These effects can be associated to  the large density of initial-state partons in the nucleus. The proton-nucleus reactions allow to investigate the different QCD dynamics at low $x$ and high gluon densities \cite{salgado} and  offer a baseline for the analyses in heavy-ion collisions. In the high energy regime, where the processes are dominated by gluons, the nucleus can be described in terms of the Color Glass Condensate (CGC) effective theory \cite{tribedy,albacete,rezaeian} as a saturated gluonic system. Then, heavy meson production may be useful to disentangle between the scenarios based on the distinct QCD descriptions of nonlinear saturation \cite{fujii,ducloue} or collinear factorization. Cold nuclear matter effects are related to the initial-state effects and constraints may be evaluated by investigating  the $D$-meson production in $pPb$ collisions.

In particular, the $D$-meson production at small-$x$ can be described within the color dipole formalism \cite{nik}. In such an approach, the phenomenology is based on the universal dipole cross section fitted to DIS data. The corresponding phenomenology has been proven suitable to evaluate inclusive and exclusive processes in the high energy limit. Moreover, a scaling property related to the DIS process at small-$x$ is naturally addressed in the parton saturation framework. Namely, the geometric scaling phenomenon is traced out on the scaling property of the dipole-target scattering amplitude, $N_{dip}(x,r)\rightarrow N_{dip}(rQ_s(x))$. The dipole cross section accounts for the nonlinear behavior and high-order corrections of QCD dynamics \cite{rauf}. The process is described in terms of a projectile that emits a gluon, which fluctuates into a quark-antiquark color dipole with definite transverse separation  that interacts with the color field of the target. Namely, the associated hard process is pictured in terms of $q\bar{q}$ dipole scattering off the target. The dipole amplitude is connected to the intrinsic dipole $k_T$-distribution, i.e., the transverse momentum distribution (TMD). In the region of large gluon transverse momentum the dipole TMD corresponds approximately to the UGD. 

In this work, we implement analytical expressions for the TMDs based on gluon saturation physics in order to obtain the double differential cross section. Predictions for the $D$-meson production in $pA$ collisions at the CERN-LHC regime are provided and a wide range of transverse momenta and rapidity are covered by the present analysis. The present work is an extension of studies performed for proton-proton ($pp$) collisions in Ref.~\cite{SampaiodosSantos:2021tfh}. We compute the $D^0$, $D^+$, and $D^{*+}$ production cross sections, the ratios of $\sigma (D^+)/\sigma (D^0)$ and $\sigma (D^{*+})/\sigma (D^0)$. Moreover,  the nuclear modification factor in $pPb$ collisions are evaluated focusing on the kinematic range available at the CERN-LHC.

The paper is organized as follows. In Sec.~\ref{form} the theoretical approach is presented, including the main expressions used in the calculations for the $D$-meson production in the color dipole framework in transverse momentum representation. In Sec.~\ref{res}, the predictions are shown for applying distinct analytical models for the gluon TMD, which are compared to the measurements obtained at the CERN-LHC by the ALICE and LHCb Collaborations. The last section summarizes our main conclusions and remarks. 

\section{Theoretical formalism}
\label{form}

Considering the $k_T$-factorization framework, the connection between the gluon UGD, ${\cal F}(x,k_{\perp}^2)$, and the usual collinear gluon distribution, $F_{g}(x,\mu_{F}^2)$, is as follows,
\begin{eqnarray}
F_{g}(x,\mu_{F}^2) = \int^{\mu^2_F}\frac{dk_{\perp}^2}{k_{\perp}^2}{\cal{F}}(x,k_{\perp}^2)\,.
\label{fg}
\end{eqnarray}
In the scenario where the momentum of the gluon in the target is particularly large, satisfying $\kappa_\perp \gg \Lambda_{\rm QCD}$, the intrinsic dipole TMD, ${\cal T}_{\mathrm{dip}}(x,k_{\perp}^2)$, is approximately equivalent to the UGD function times $\alpha_s$ \cite{gbw, bart, albcgc, altcgc}. This indicates that a connection between the $k_\perp$-factorization and the dipole framework is feasible, i.e.,
\begin{eqnarray} 
{\cal T}_{\mathrm{dip}}(x,k_{\perp}^2) \simeq \alpha_s\,{\cal F}(x,k_{\perp}^2)\,.
\label{dip_kt}
\end{eqnarray}
This approximated relation can be safely employed on the evaluation of the $D$-meson production. In general, the  gluon UGD is not well determined at small $k_{\perp}$, however there is a correspondence concerning the intrinsic dipole TMD and the color dipole cross section $\sigma_{q\bar{q}}$ (see, e.g. Refs.~\cite{nik1,bart}).
Therefore, for a given dipole cross section model is possible to determine the respective TMD by taking a specific Fourier transform.

The QCD dipole formalism assumes that the production process is described via a color dipole that interacts with the color field of the nucleon/nucleus considering the target rest frame. The $D$-meson production is determined by the cross section of the process $g+N\rightarrow Q\bar{Q}+X$, where the $Q\bar{Q}$ produced in singlet and color-octet states comes from a virtual gluon fluctuation. Consequently, the cross section associated to the hadronic collision $pp \rightarrow Q\bar{Q}X$ is given by
\begin{eqnarray} 
\frac{d^{4}\sigma{(pp \rightarrow Q\bar{Q}X)}}{dy d\alpha d^2p_T} = F_{g}(x_1,\mu_{F}^2)\,
\frac{d^{3}\sigma{(gp \rightarrow Q\bar{Q}X)}}{d\alpha d^2p_T} \,, 
\label{hdeq}
\end{eqnarray}
where there is a convolution between the $gp \rightarrow Q\bar{Q}X$ cross section and the projectile gluon UGD. Along with this, the form as the cross section is obtained in Eq.~(\ref{hdeq}) is similar to that of the $k_T$-factorization scenario. Moreover, in Eq.~(\ref{hdeq}) $y$ is the rapidity and $p_T$ is the transverse momentum of the heavy quark, while $\alpha$ 
($\bar{\alpha} = 1 - \alpha$) stands for the fractional gluon momentum exchanged with the heavy quark (antiquark). In the momentum representation the heavy quark TMD is written in connection with the dipole TMD \cite{vic},
\begin{eqnarray} 
\frac{d^3\sigma{(gp \rightarrow Q\bar{Q}X)}}{d\alpha d^2 p_T } &=& 
\frac{1}{6\pi} \int \frac{d^2 \kappa_{\perp}}{\kappa^{4}_{\perp}}  \alpha_s(\mu_{F}^2)\, {\cal T}_{\mathrm{dip}}(x_2,\kappa^{2}_{\perp})\,\bigg\{\bigg[\frac{9}{8}{\cal{I}}_0(\alpha,\bar{\alpha},p_T) - \frac{9}{4} {\cal{I}}_1(\alpha,\bar{\alpha},\vec{p}_T,\vec{\kappa}_{\perp}) \nonumber \\ 
&+& {\cal{I}}_2(\alpha,\bar{\alpha},\vec{p}_T,\vec{\kappa}_{\perp}) + \frac{1}{8}{\cal{I}}_3(\alpha,\bar{\alpha},\vec{p}_T,\vec{\kappa}_{\perp})\bigg] + \left[\alpha \longleftrightarrow \bar{\alpha}\right]\bigg\} \,,  
\label{proxs} 
\end{eqnarray}
with $\alpha_s(\mu_{F}^2)$ being the running coupling in the one-loop approximation evaluated at the scale $\mu_{F}^2 = M^{2}_{Q\bar{Q}}$, where $M_{Q\bar{Q}}$ is the invariant mass of the $Q\bar{Q}$ pair, $
M_{Q\bar{Q}}\simeq 2\sqrt{m_{Q}^2+p_T^2}$ ($m_Q$ is the heavy quark mass). Furthermore, $x_{1}$ and $x_{2}$ correspond to the fractional light-cone momentum of the projectile and target given by $x_{1,2} = \frac{M_{Q\bar{Q}}}{\sqrt{s}}\, e^{\pm y}$, with $\sqrt{s}$ denoting the collision center-of-mass energy. Additionally, the Eq.~(\ref{proxs}) contains the auxiliary quantities ${\cal{I}}_i(\alpha,\bar{\alpha},p_T)$ ($i=0,1,2,3$) -- we quote Refs.~\cite{SampaiodosSantos:2021tfh,vic} for their corresponding expressions. 

Since that the UGD is not obtained from the first principles, it requires modeling, and  distinct parameterizations for the UGDs are found in the literature. In this work, we will consider the analytical parametrizations from Refs.~\cite{gbw,mpm} for the UGD in protons. The following UGD parameterization is resulting from the Golec-Biernat and W\"usthoff (GBW) dipole cross section \cite{gbw,gbw1} based on the gluon saturation assumption, assuming the form
\begin{eqnarray}
{\cal F}_{GBW}(x,k_{\perp}^2)=\frac{3\,\sigma_{0}}{4 \pi^2\alpha_{s}} \frac{k_{\perp}^4}{Q_{s}^2}\,\mathrm{exp}\left(-\frac{k_{\perp}^2}{Q_{s}^2}\right) \,,
\label{FGBW}
\end{eqnarray}
with $\alpha_{s} = 0.2$ and $Q_{s}^2(x) = (x_0/x)^{\lambda}$~GeV${^2}$ being the saturation scale in the proton. The parameters $\sigma_{0}$, $x_{0}$, and $\lambda$ are extracted from a fit to the proton structure function $F_2$, which was recently done in Ref.~\cite{gbwfit}.

On the other hand, an approach that accounts for the geometric scaling present in charged hadron production in $pp$ collisions combined with a Tsallis-like distribution extracted from the measured hadron spectrum is proposed in Ref.~\cite{mpm} (hereafter MPM model). The corresponding UGD is expressed as:
\begin{eqnarray}
{\cal F}_{MPM}(x, k_{\perp}^2)=\frac{3\,\sigma_{0}}{4\pi^2\alpha_{s}}\frac{(1+\delta n)}{Q_{s}^2}
\frac{k_{\perp}^4}{\left(1+\frac{k_{\perp}^2}{Q_{s}^2} \right)^{(2+\delta n)}}\, ,
\label{FMPM}
\end{eqnarray}
with $\alpha_{s} = 0.2$ and $Q_{s}^2(x) = (x_0/x)^{0.33}$~GeV$^2$. The powerlike behavior of the gluons produced at high momentum spectrum is determined via the function $\delta n = a\tau^{b}$, where $\tau$ is the scaling variable defined as $\tau = k_{T}^2/Q_{s}^2$. Moreover, the set of parameters $\sigma_{0}$, $x_{0}$, $a$, and $b$ are fitted from DIS data available at small-$x$. Thus, we consider the parameters from Fit A in Ref.~\cite{mpm} for our calculuations.

An essential ingredient to obtain the differential distribution of $D$-meson is to account for the hadronization process of the heavy quarks. Hence, the differential distribution of open heavy mesons is given by the convolution of the heavy quark cross section and the fragmentation function,
\begin{eqnarray} 
\frac{d^{3}\sigma{(pp \rightarrow DX)}}{dY d^2P_T} = \int_{z_{\mathrm{min}}}^1 \frac{dz}{z^2} 
D_{Q/D} (z,\mu_{F}^2) \int_{\alpha_{\mathrm{min}}}^1 d\alpha \frac{d^{4}\sigma{(pp \rightarrow Q\bar{Q}X)}}{dyd\alpha d^2p_T} \,,
\label{Dmes}
\end{eqnarray}
where the heavy quark light-cone momentum exchanged with the $D$-meson is denoted by $z$. In addition, $D_{Q/D}(z,\mu_{F}^2)$ represents the fragmentation function. In the calculation the KKKS parametrization \cite{kkks08} will be considered. Furthermore, the mass, rapidity, and transverse momentum of the $D$-meson are given,  namely $m_D$, $Y=y$ , and $P_T$, respectively \cite{maciula}. The quark and charmed hadron transverse momenta are related by $p_{T} = P_{T}/z$. The lower limits regarding the $z$ and $\alpha$ integration are defined by $z_{\mathrm{min}}= (m_{\perp}/\sqrt{s}) e^{Y}$ and $\alpha_{\mathrm{min}}=(z_{\mathrm{min}}/z)\sqrt{(m_{Q}^2 z^2 + P_{T}^2)/m_{\perp}^2}$, respectively. The $D$-meson transverse mass is given by $m_{\perp} = \sqrt{m_{D}^2 + P_{T}^2}$.

As shown in Ref.~\cite{SampaiodosSantos:2021tfh} an approximate expression for the $p_T$-spectrum in $pp$ collisions can be evaluated by means of the GBW parameterization. The kinematic domain established here implies that the hard scale $\mu_F$ achieves higher values than the saturation scale, $\mu_F^2/Q^{2}_s(x)\gg 1$, and in this limit one has $F_g^{GBW}\approx 3\sigma_0Q^{2}_s(x_1)/(2\pi)^2\alpha_s$. It can be shown that the $D$-meson spectra is given approximately by \cite{SampaiodosSantos:2021tfh}:
\begin{eqnarray}
\frac{d^{3}\sigma{(pp \rightarrow DX)}}{dY d^2P_T} \approx \left[\frac{\langle z\rangle \,\sigma_0 }{2(2\pi)^2}\right]^2\frac{Q_s^2(x_1)Q_s^2(x_2)}{5}\left[ \frac{9m_c^4\langle z\rangle^4 +25m_c^2\langle z\rangle^2P_T^2 + 9P_T^4}{(m_c^2\langle z\rangle^2+P_T^2)^4}\right]\,,
\label{approximation}
\end{eqnarray}
where $\langle z\rangle_c $ is the average momentum fraction \cite{kkks08},
\begin{eqnarray}
\langle z\rangle_c (\mu_F) = \frac{\int_{z_{cut}}^{1}dz z D_c(z,\mu_F)}{\int_{z_{cut}}^{1}dz D_c(z,\mu_F)}\,, 
\end{eqnarray}
being $B_c$ the branching fraction $c \to D$ and $x_{cut}=0.1$ \cite{kkks08}.  We will assume $\langle z\rangle \equiv \langle z\rangle_c (2m_c)$. Here we employ the KKKS fragmentation function that considers $\langle z\rangle_c (\mu_F=2m_c) =$ 0.573, 0.571, and $0.617$ for $D^0$, $D^+$ and $D^{*+}$, respectively.

Here we focus on the evaluation of the $D$-meson production spectrum in proton-nucleus ($pA$) collisions. Considering a heavy target colliding at high-energy regime, the nuclear QCD effects are present and, in particular, those associated to multiple parton scattering and nonlinear gluon saturation.
Within the color dipole approach, such nuclear effects can be embedded onto the dipole-nucleus amplitude $N_A$ by means of the geometric scaling (GS) property derived from parton saturation models \cite{salgado1}. The geometric scaling (GS) is based on the assumption that the nuclear effects are absorbed into the saturation scale and on the transverse area of the colliding nucleus, establishing an $A$-dependence in the scattering cross section. Thus, the nuclear effects are embedded into the saturation scale and on the nucleus transverse area, $S_A=\pi R_A^2$, in correlation to the proton case, $S_p=\sigma_0/2=\pi R_p^2$, where the nucleus radius is given by $R_A \simeq 1.12 A^{1/3}$.
Therefore, it is necessary to replace the proton saturation scale by the corresponding nuclear saturation scale, consequently,
\begin{eqnarray}
Q_{s,A}^2&=&Q_{s,p}^2\left(\frac{A\pi R_p^2}{\pi R_A^2}\right)^{\Delta},  \label{qs2A} \\
N_A(x,r) & = & N(rQ_{s,p}\rightarrow rQ_{s,A})\,,
\label{qs2AN}
\end{eqnarray}
where the parameters $\Delta = 1/\delta$ and $S_p=\pi R_p^2$ are adjusted by data producing $\delta = 0.79$ and $\pi R_p^2=1.55$~fm$^2$ \cite{salgado1}.
Hence, the assumptions encoded in the GS are converted into the $D$-meson cross section which is appropriately rescaled  in the following form, 
\begin{eqnarray}
\frac{d^{3}\sigma{(pA \rightarrow DX)}}{dY d^2P_T} = \left(\frac{S_A}{S_p}\right) \left.\frac{d^{3}\sigma{(pp \rightarrow DX)}}{dY d^2P_T}\right|_{Q_{s,p}^2(x_2)\rightarrow Q_{s,A}^2(x_2)}\,.
\label{prescr}
\end{eqnarray}
The approximation above has been tested against the experimental measurements for prompt photon production in $pA$ and $AA$ collisions in Refs. \cite{SampaiodosSantos:2020lte,SampaiodosSantos:2020raq,Santos:2020kgv}.

Using the approximate analytical expression for open heavy meson production off protons, Eq. (\ref{approximation}), and the GS arguments presented above, the following parametrization for the nuclear modification factor is obtained in the limit $P_T\rightarrow \infty$:
\begin{eqnarray}
R_{pA}(y,P_T) =   \frac{d^{3}\sigma(pA \rightarrow DX)/dY d^2P_T}{A\, d^{3}\sigma (pp \rightarrow DX)/dY d^2P_T}\approx \left(\frac{S_A}{AS_p}  \right)\frac{Q_{s,A}^2(x_2)}{Q_{s,p}^2(x_2)} =\left(\frac{AS_p}{S_A}  \right)^{\Delta -1}\,,
\end{eqnarray}
where the small-$x$ data on $\gamma^*A$ collisions support an
increase of $Q_{s,A}^2$  stronger than $A^{1/3}$ since $\Delta \simeq 1.27$. Hence, at sufficiently large $p_T$ one expects $R_{pA}\simeq 1.26$. In case $\Delta = 1$ which corresponds to $Q_{s,A}^2\propto A^{1/3}$ one has $R_{pA}=1$.  The low-$P_T$ limit of the nuclear modification factor has to be determined numerically. However, the qualitative behavior should be similar to the ratio for gluon production in $pA$ in the context of CGC approach. Namely, $R_{pA}(k_{\perp}, y)\approx k_{\perp}^2/Q_{s,A}^2\ln (k_{\perp}^2/\Lambda^2) $ for $k_{\perp}^2<Q_{s,A}^2$, with $\Lambda$ being some infrared cutoff \cite{Kharzeev:2003wz,Jalilian-Marian:2005ccm}.

Alternatively, the $D$-meson production in $pA$ collisions can be computed by using the nuclear version for the unintegrated gluon distribution in Eq. (\ref{proxs}). Namely, the UGD of the proton is substituted for the nucleus one, ${\cal F}_{A}$. Here we will apply the model for ${\cal F}_{A}$ provided in Ref. \cite{armesto}, which is based in a Glauber-Gribov expression for the dipole-nucleus cross section, $\sigma_{dA}(x,r,b)$ at a given impact parameter $b$. Considering a given $b$ and a Bessel-Fourier transform, one can associate the ${\cal F}_{A}$ with the dipole-nucleus cross section as follows \cite{armesto1,armesto2}:
\begin{eqnarray}
{\cal F}_{A}(x,r,b) = - \frac{3k^{2}_{\perp}}{4\pi^2\alpha_{s}} \int \frac{d^{2}r}{2\pi}\,e^{i\vec{k_{\perp}} \cdot\,\vec{r}}\, \sigma_{dA}(x,r,b),\quad \sigma_{dA}(x,r,b)= 2\left[1-\exp{\left(-\frac{1}{2}AT_A(b)\sigma_{q\bar{q}}(x,r)\right)}\right]\nonumber \\
\label{fA}
\end{eqnarray}
and by applying the technique described in Ref. \cite{armesto2} the UGD for the nucleus (using the GBW parametrization for $\sigma_{q\bar{q}}(x,r)$) can be written as:
\begin{eqnarray}
{\cal F}_{A}(x,r,b) = \frac{3}{\pi^2\alpha_{s}}\frac{k^{2}_{\perp}}{Q_{s}^2} \sum_{n=1}^{\infty} \frac{(-B)^{n}}{n!} \sum_{\ell=0}^{n} C_{n}^{\ell}\frac{(-1)^\ell}{\ell} \,\mathrm{exp}\left(-\frac{k_{\perp}^2}{\ell\,Q_{s}^2}\right)\,,
\label{ugdfnuc}
\end{eqnarray}
where $B = AT_{A}(b)\sigma_{0}/2$ and $T_{A}(b)$ represents the nuclear profile function. For large nucleus the series is fastly convergent and  in our calculations we take $n=7$. Hereafter we will refer to this model by UGDnuc.

Based on the expressions introduced before to calculate the $D$-meson $P_T$ and $Y$ distributions in $pA$ collisions, in the next section we employ the referred UGD parameterizations in order to obtain the corresponding predictions and comparing them with the experimental measurements reported by the LHC collaborations.

\section{Results and discussions}
\label{res}

\begin{figure*}[!t]
\begin{tabular}{cc}
\includegraphics[width=0.5\textwidth]{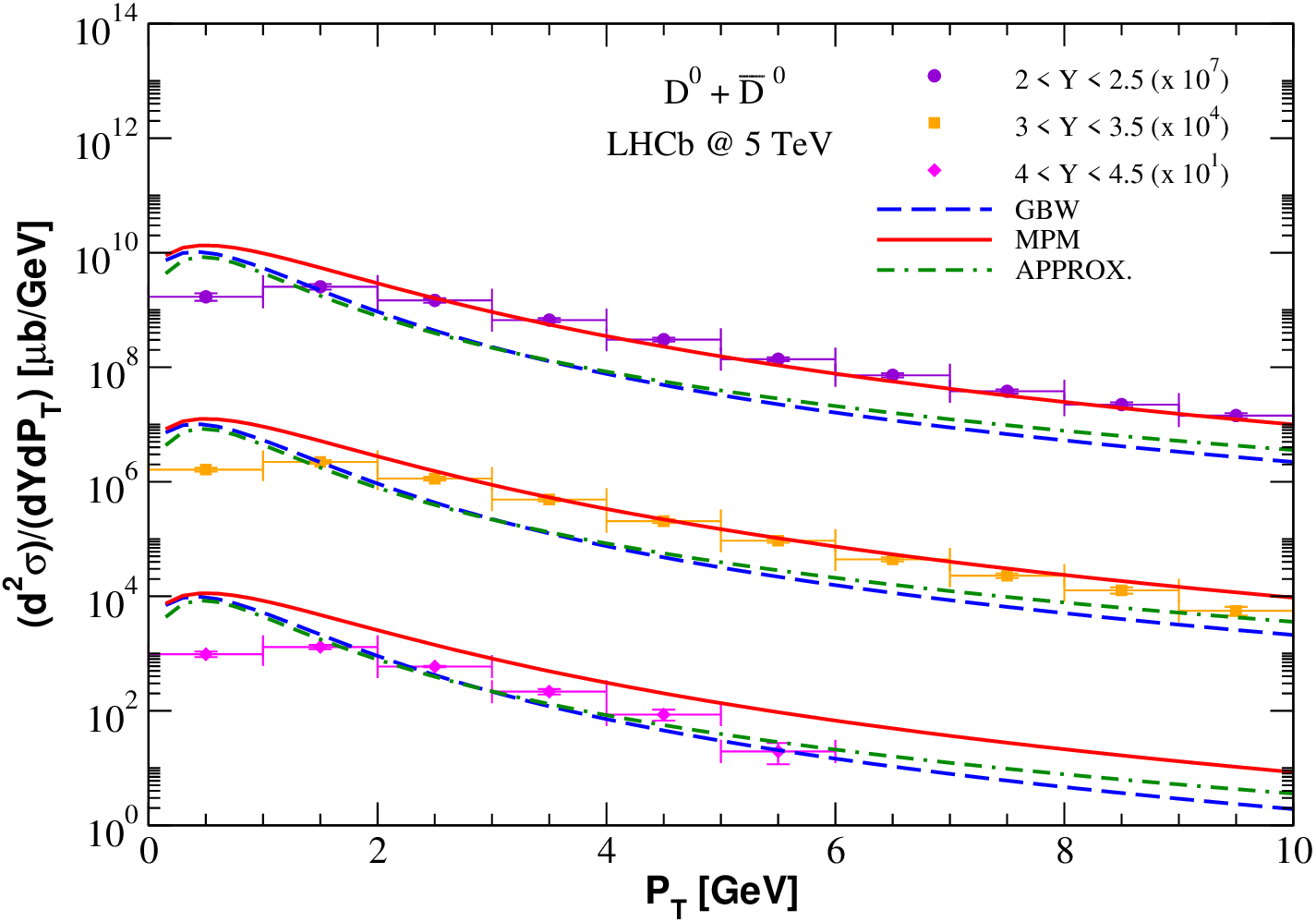} & \includegraphics[width=0.5\textwidth]{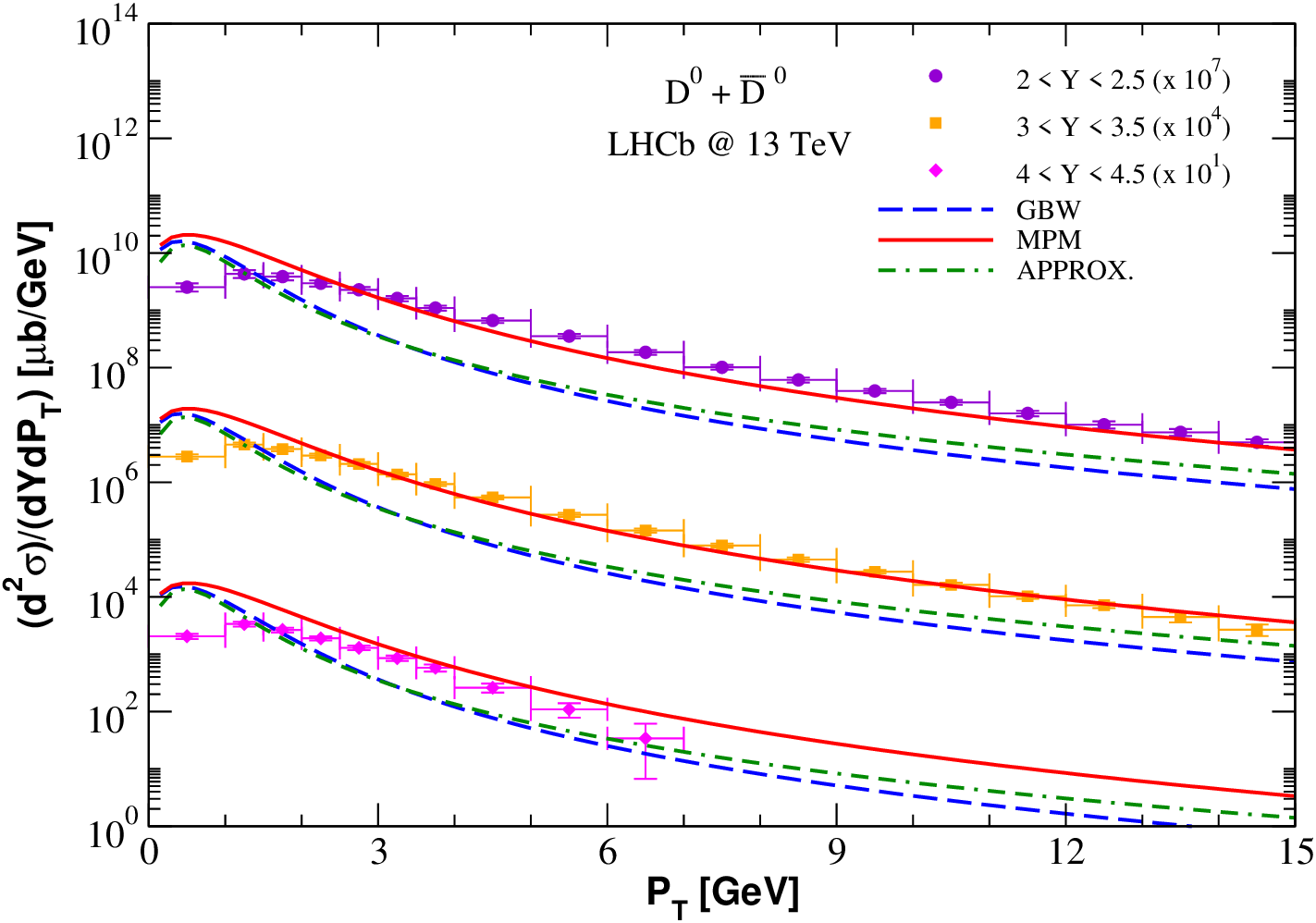} 
\end{tabular}
\caption{The differential $D^0$ production cross section as a function of $P_T$ and $Y$ in $pp$ collisions at $\sqrt{s} = 5$ and $13$ TeV considering forward rapidity bins. The results with the GBW, MPM and APPROX. models are directly compared against the experimental measurements provided by the LHCb Collaboration \cite{LHCbpp5, LHCbpp13}.}
\label{pp5_13}
\end{figure*}

Before investigating the $D$-meson production in nuclear collisions, we show the theoretical predictions that can be found in Ref.~\cite{SampaiodosSantos:2021tfh} compared with the current setup with LHC data in nucleon-nucleon collisions \cite{LHCbpp5, LHCbpp13}. Comparing the LHC $pp$ data on the $P_T$ spectrum of $D$-meson and theoretical calculations is important for  consistency and applicability of the calculations within the color dipole approach. The corresponding predictions regarding the $D^0$ production at center of mass energy of 5 and 13~TeV are presented in Fig.~\ref{pp5_13} and they demonstrate that the color dipole approach works -- for more details, see Ref.~\cite{SampaiodosSantos:2021tfh}.

\begin{figure*}[!t]
\begin{tabular}{cc}
\includegraphics[width=0.5\textwidth]{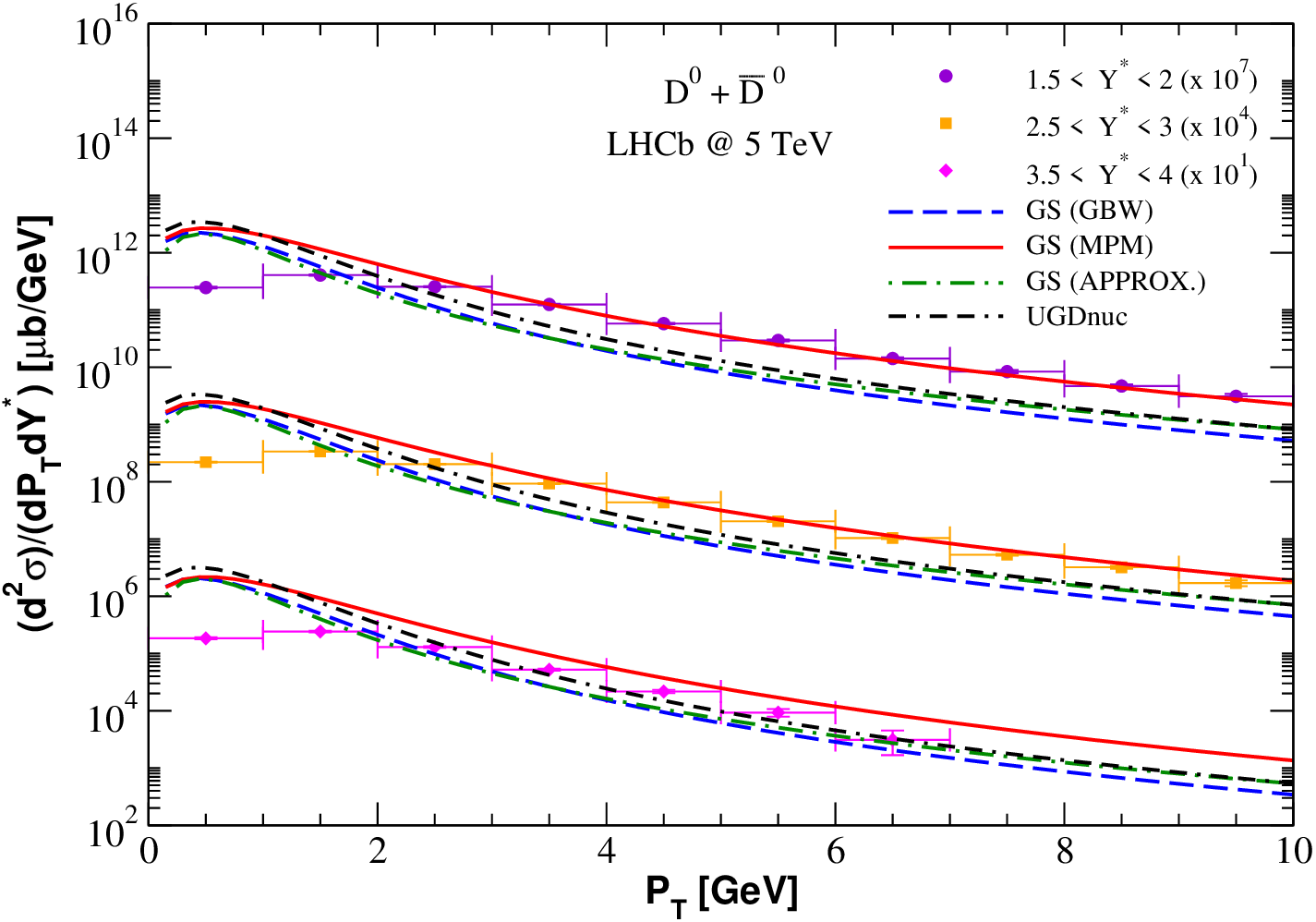} & \includegraphics[width=0.5\textwidth]{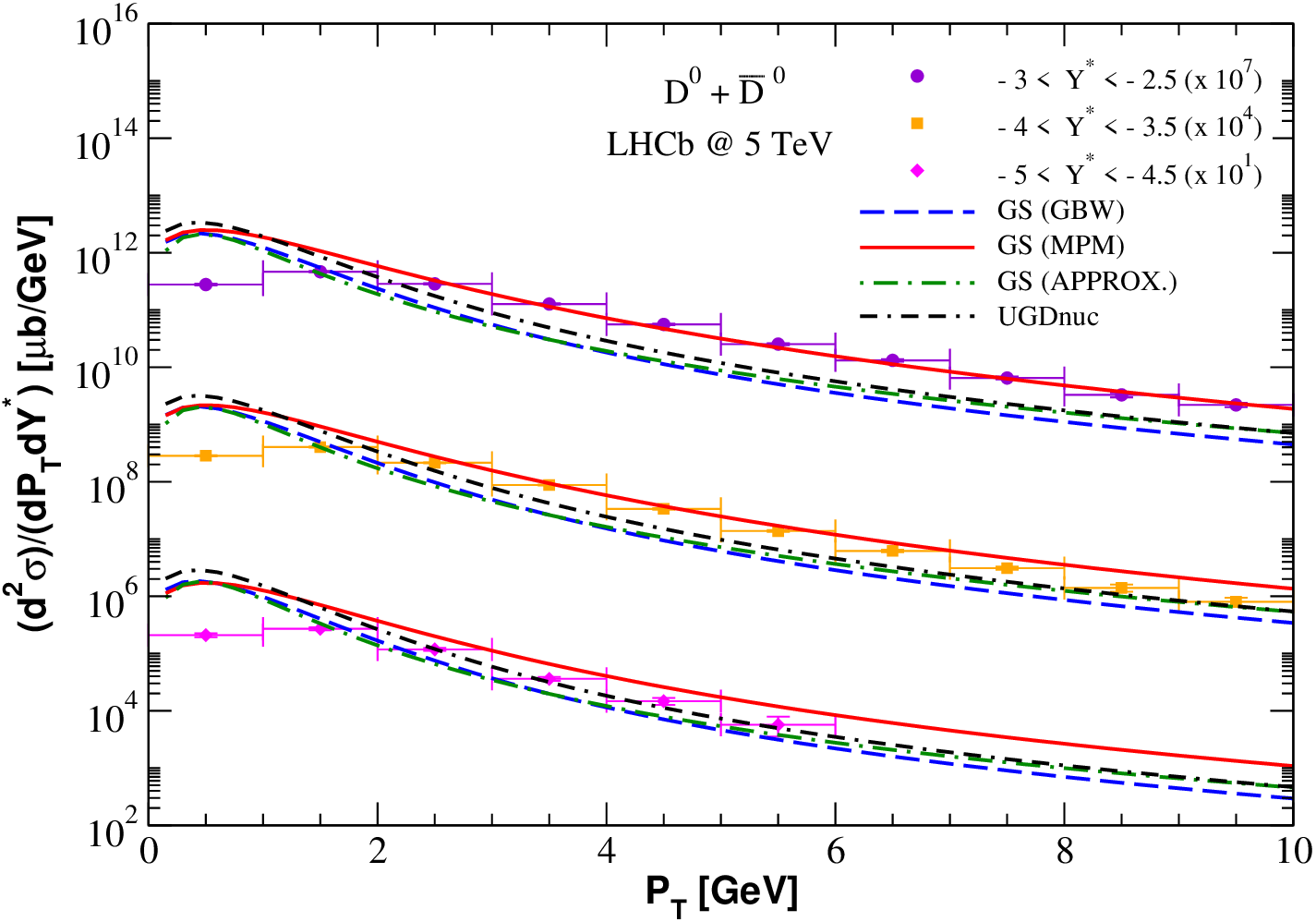} 
\end{tabular}
\caption{The differential $D^0$ production cross section as a function of $P_T$ and $Y$ in $pPb$ collisions at $\sqrt{s} = 5$ TeV considering three forward and backward rapidity bins. The predictions given by the GS (GBW), GS (MPM), GS(APPROX.) and UGDnuc models are directly compared with the experimental data from the LHCb Collaboration \cite{LHCb5}.}
\label{pPb5}
\end{figure*}

Our analyzes correspond to the $D$-meson production in $pA$ collisions in terms of the transverse momentum and center-of-mass rapidity where a comparison to the experimental measurements provided by ALICE and LHCb Collaborations is done. The predictions  are obtained considering the color dipole framework in transverse momentum representation with different unintegrated gluon distributions (UGDs) and by applying the GS property. The results with the UGDs GBW and MPM take into account the GS phenomenon following the prescription in Eq.~(\ref{prescr}). The results will be referred hereafter as \emph{GS (GBW)} and \emph{GS (MPM)}. Moreover, we provided results by employing the nuclear UGD presented in Eq.~(\ref{fA}) denoted as \emph{UGDnuc}. Finally, the approximated expression given in Eq.~(\ref{approximation}) is labeled as \emph{GS (APPROX.)}.

Let us first present the predictions for the double differential cross section for $D^0$ production including the charge conjugated states in $pPb$ collisions at $\sqrt{s} = 5$~TeV. In Fig.~\ref{pPb5} the respective results are compared to the LHCb data \cite{LHCb5} for different forward and backward rapidity bins.
Considering both forward and backward rapidities, we observe that the results with \emph{GS (MPM)} describe the data except for the rapidity intervals, $3.5 < Y^{*} < 4$ and $-5 < Y^{*} < -4.5$. However, \emph{GS (GBW)} and \emph{GS (APPROX.)} approaches reproduce similar predictions in almost the entire $P_T$-distribution. A difference between them occurs in the spectrum at $P_T > 6$~GeV. Also a reasonable agreement with the experimental data is observed in the very forward ($3.5 < Y^{*} < 4$) and very backward ($-5 < Y^{*} < -4.5$) rapidity bins. In these configurations, we can conclude that the predictions from the approximate expression mimic the estimates obtained with the \emph{GS (GBW)}.
In particular the GBW parameterization presents a Gaussian shape that takes place in Eqs.~(\ref{dip_kt}) and (\ref{FGBW}), which results in a suppression that underestimates the data.
Concerning the \emph{UGDnuc} results in forward/backward rapidity bins the model provides a reasonable description of the experimental points in  a narrow $P_T$ interval ($2 < P_T < 3$~GeV). On the other hand, a better adjustment to the experimental measurements is found at the rapidity regions $3.5 < Y^{*} < 4$ and $-5 < Y^{*} < -4.5$.
It should be noticed  that the results with the \emph{UGDnuc}, where the parameter $B < 3$, are reliable [see Eq.~(\ref{ugdfnuc})] due to the fast series convergence.

\begin{figure*}[!t]
\begin{tabular}{cc}
\includegraphics[width=0.5\textwidth]{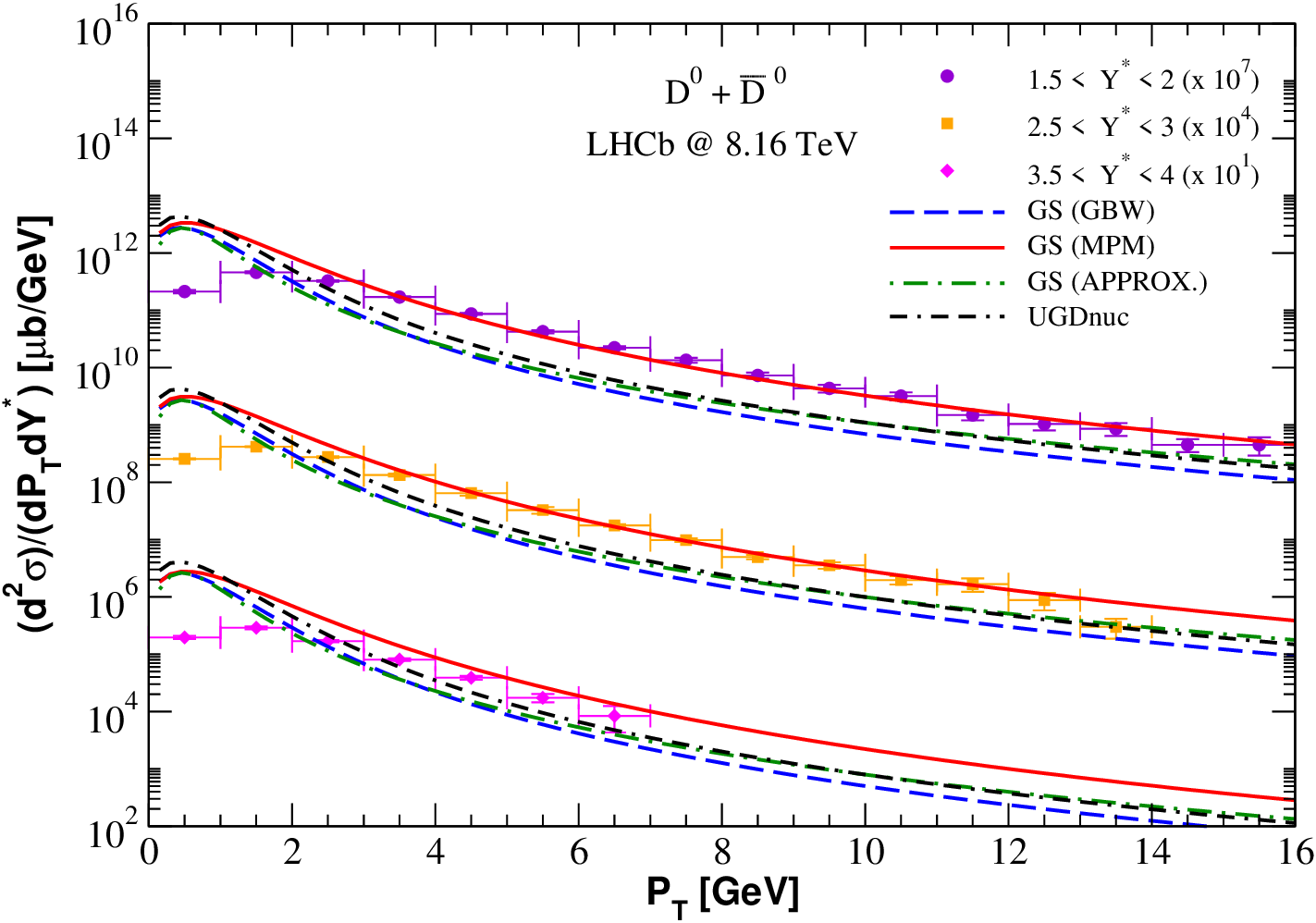} & \includegraphics[width=0.5\textwidth]{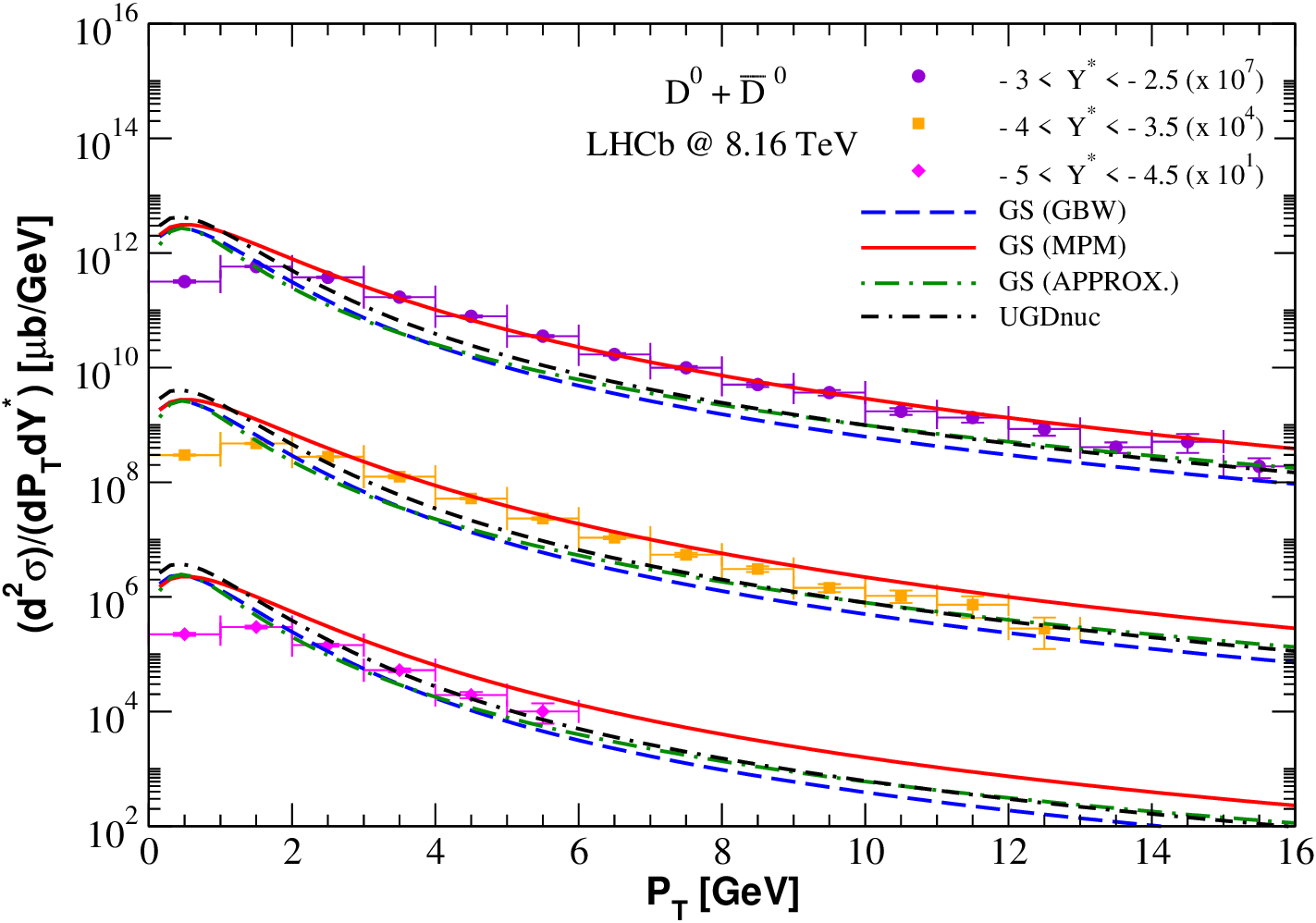} 
\end{tabular}
\caption{The differential $D^0$ production cross section as a function of $P_T$ and $Y$ in $pPb$ collisions at $\sqrt{s} = 8.16$~TeV considering three forward and backward rapidity bins. The predictions given by the \emph{GS (GBW)}, \emph{GS (MPM)}, \emph{GS (APPROX.)}, and \emph{UGDnuc} models are directly compared with the preliminary experimental data from the LHCb Collaboration \cite{LHCb816}.}
\label{pPb816}
\end{figure*}

The cross section for $D^0$ + $\bar{D}^0$ production in $pPb$ collisions at $8.16$~TeV is shown in Fig.~\ref{pPb816} for forward/backward rapidity bins. Comparison is done with the  preliminary measurements reported by the LHCb Collaboration \cite{LHCb816}.
Here the results indicate the same pattern found at $5$~TeV, however one has a wider $P_T$ spectrum at $\sqrt{s} = 8.16$~TeV. Also, the difference between the \emph{GS (GBW)} and \emph{GS (APPROX.)} results become more apparent for $P_T > 6$~GeV in comparison with $\sqrt{s} = 5$~TeV. At the same time, the \emph{GS (APPROX.)} and \emph{UGDnuc} models show equivalent results considering the kinematic region $P_T > 6$~GeV.

\begin{figure*}[!t]
\begin{tabular}{c}
\includegraphics[width=\textwidth]{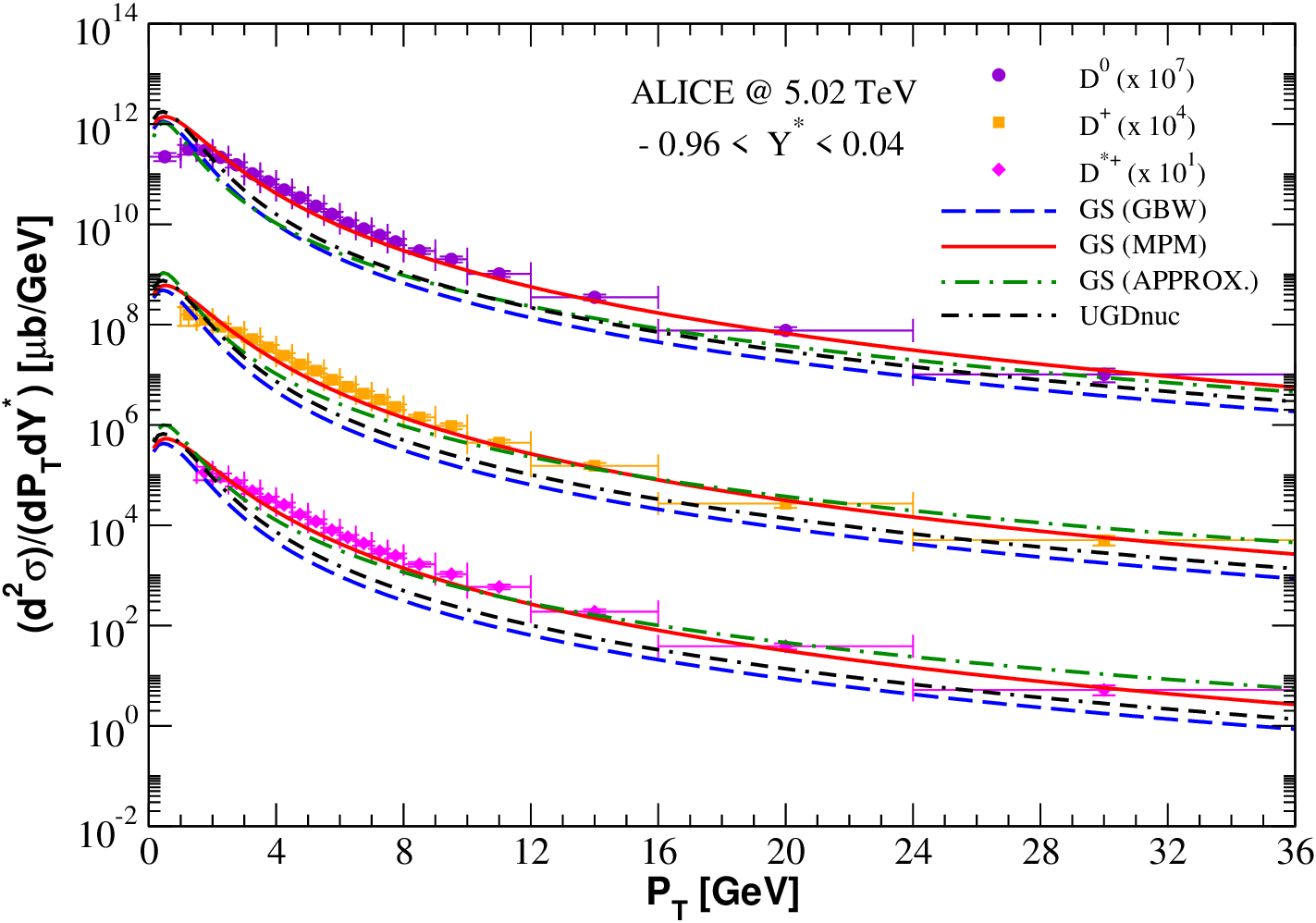} 
\end{tabular}
\caption{The differential $D^0$, $D^+$, and $D^{*+}$ production cross sections as a function of $P_T$ and $Y$ in $pPb$ collisions at $\sqrt{s} = 5.02$~TeV considering $-0.96 < Y^{*} < 0.04$. The predictions given by the \emph{GS (GBW)}, \emph{GS (MPM)}, \emph{GS(APPROX.)}, and \emph{UGDnuc} models are directly compared to the experimental data from the ALICE Collaboration \cite{ALICE502}.}
\label{pPb502}
\end{figure*}

\begin{figure*}[!t]
\begin{tabular}{cc}
\includegraphics[scale=0.35]{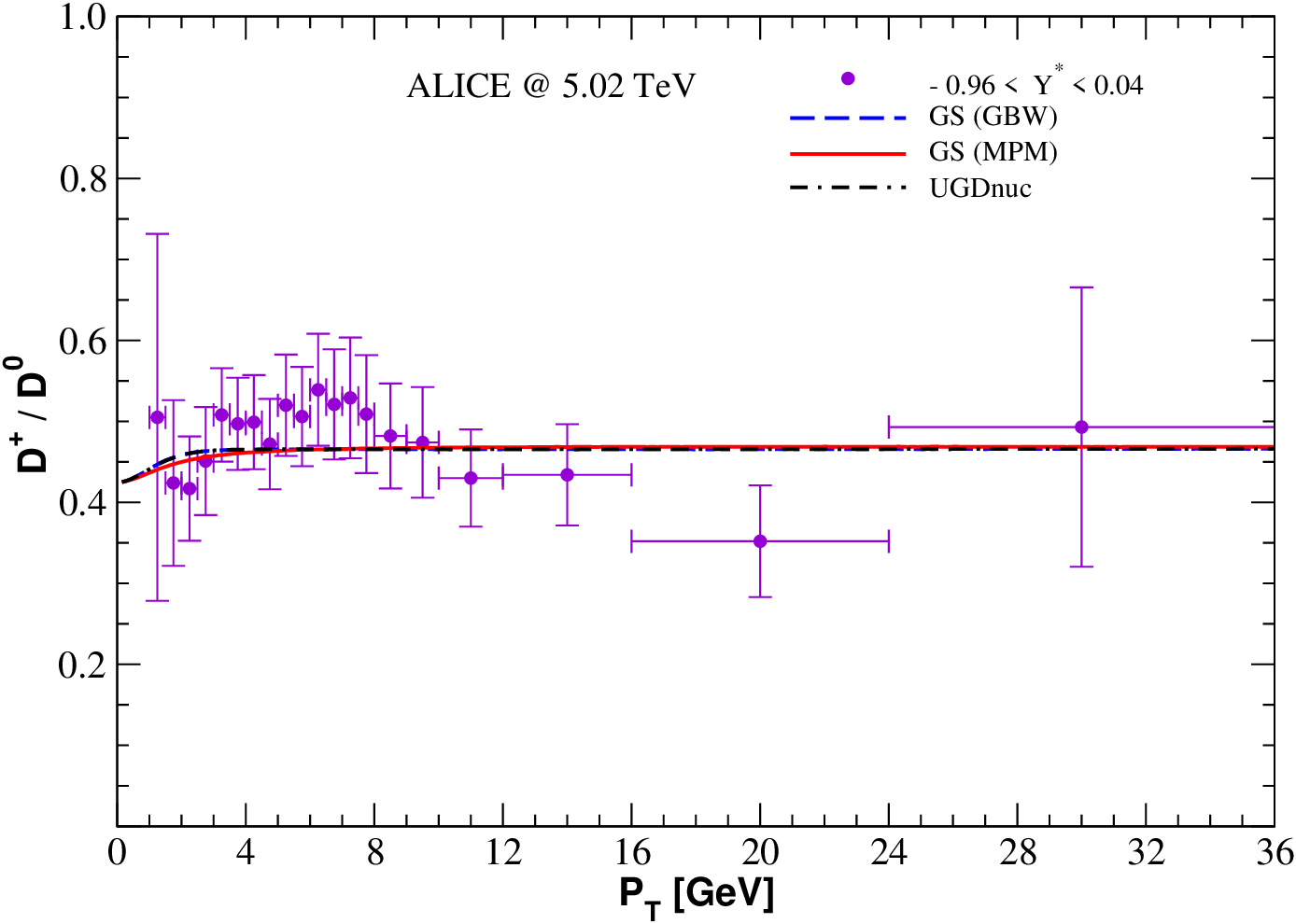} & \includegraphics[scale=0.35]{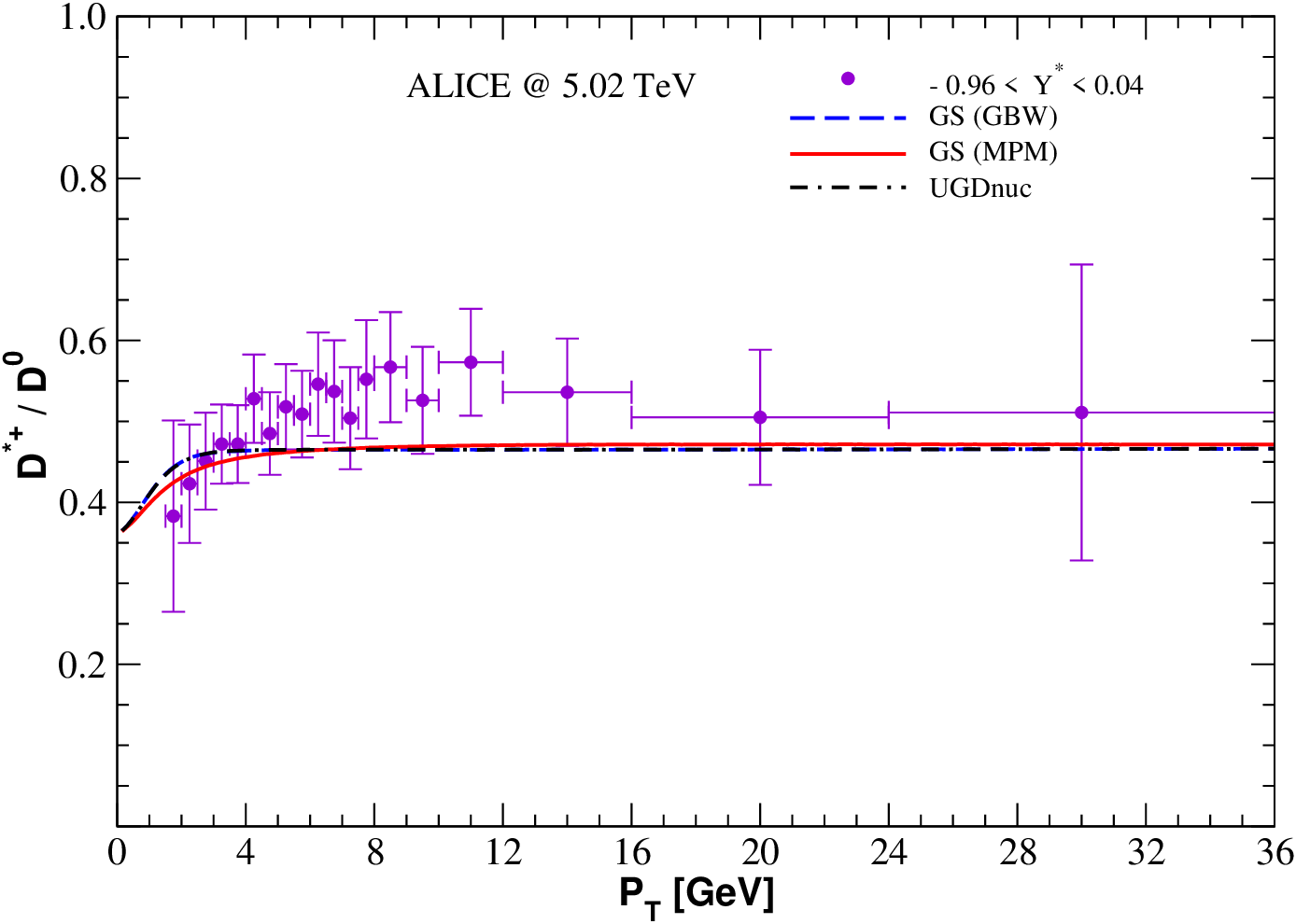}
\end{tabular}
\caption{The ratios as function of $P_T$ considering the $D^+/D^0$ (left panel) and $D^{*+}/D^0$ (right panel) differential production cross sections. The predictions given by the \emph{GS (GBW)}, \emph{GS (MPM)}, and \emph{UGDnuc} models are directly compared to the experimental data from the ALICE Collaboration \cite{ALICE502}.}
\label{ratio}
\end{figure*}

\begin{figure*}[!t]
\begin{tabular}{cc}
\includegraphics[width=0.5\textwidth]{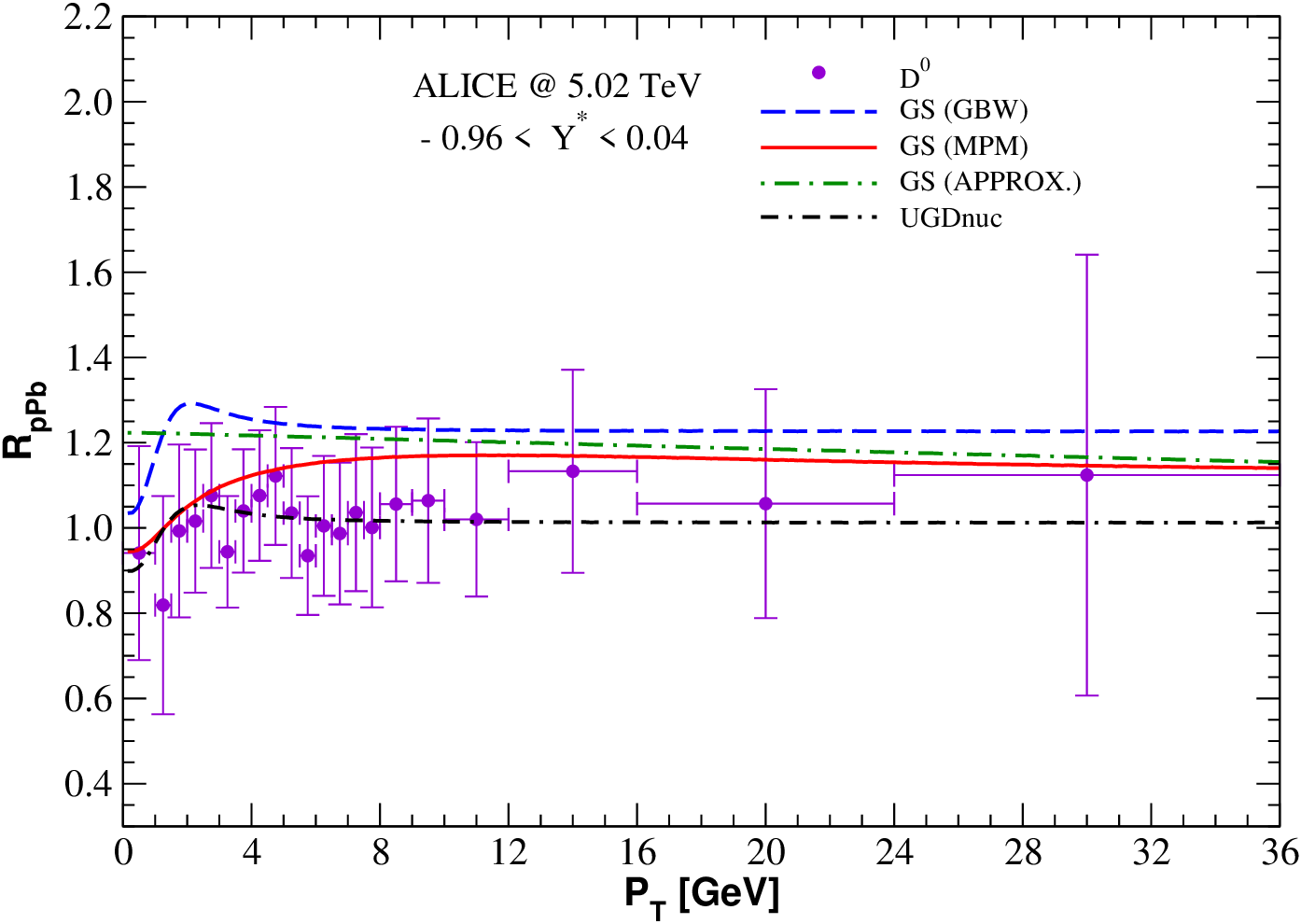}  \includegraphics[width=0.5\textwidth]{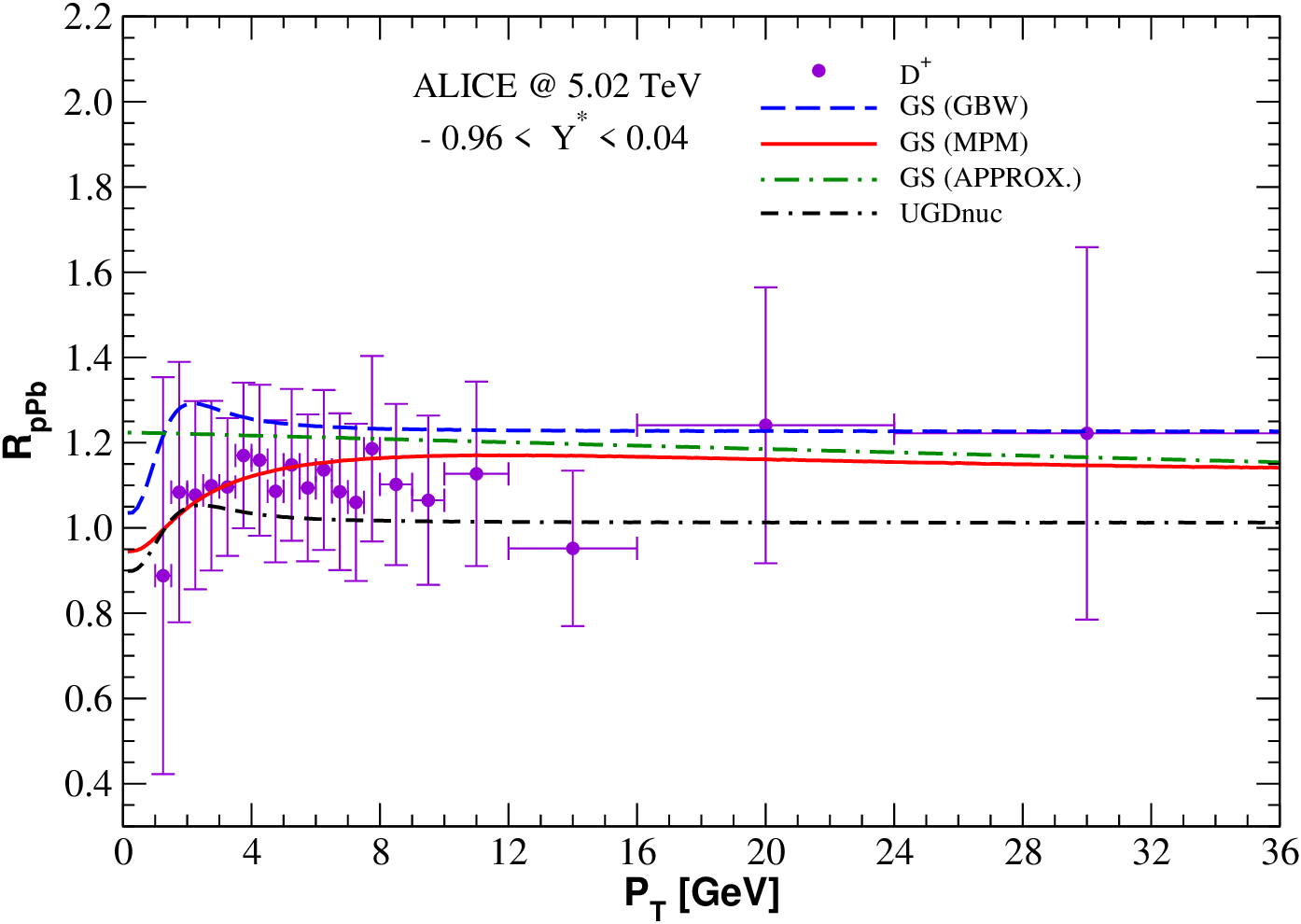} \\
\includegraphics[width=0.5\textwidth]{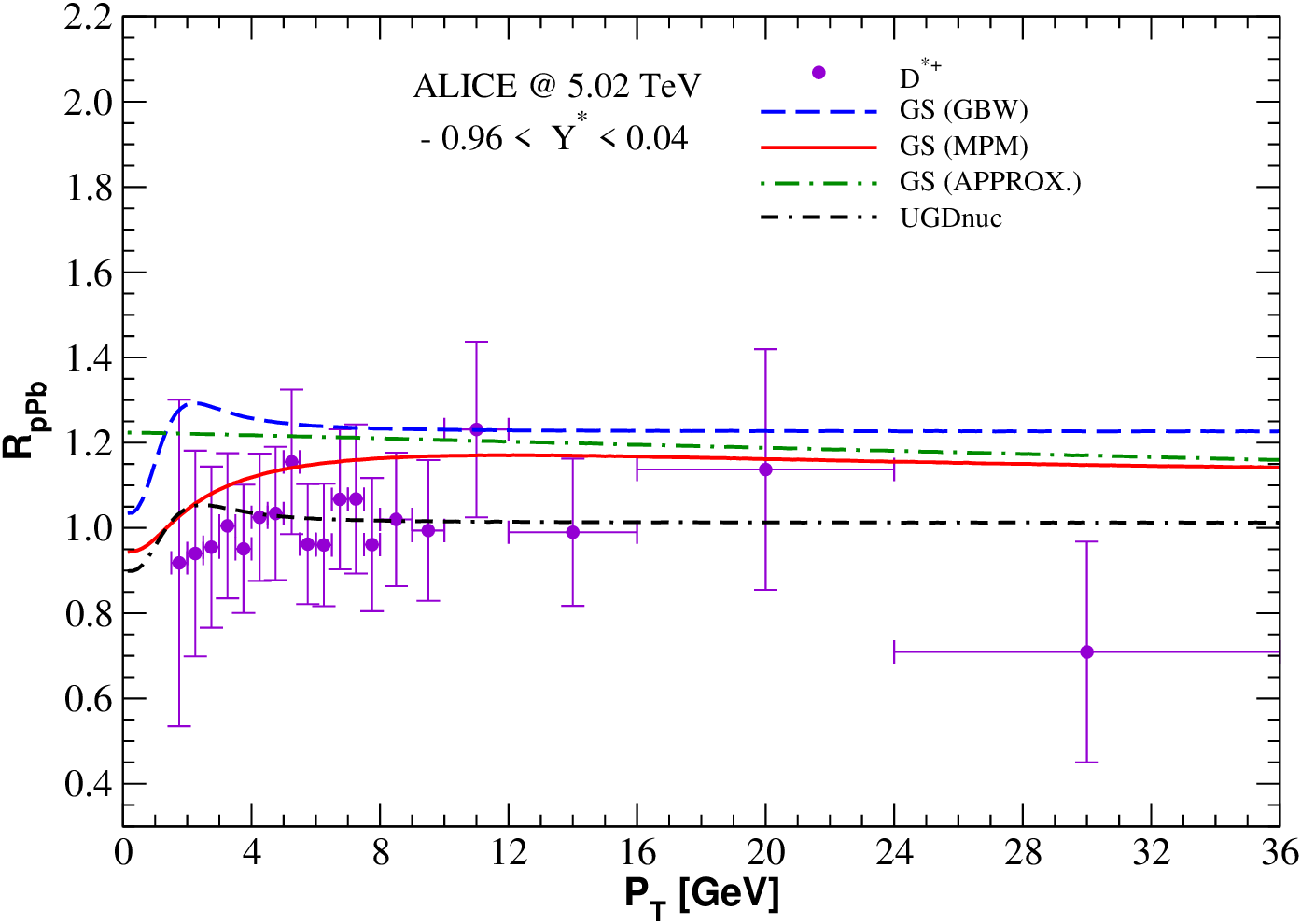}
\end{tabular}
\caption{The nuclear modification factor for $D^0$ (upper left panel), $D^+$ (upper right panel), and $D^{*+}$ (bottom panel) meson production in $pPb$ collisions at $\sqrt{s} = 5.02$~TeV considering $-0.96~<~Y^{*}~<~0.04$. The predictions given by the \emph{GS (GBW)}, \emph{GS (MPM)}, \emph{GS (APPROX.)}, and \emph{UGDnuc} models are directly compared to the experimental data from the ALICE Collaboration \cite{ALICE502}.}
\label{rpPb}
\end{figure*}

\begin{figure*}[!t]
\begin{tabular}{cc}
\includegraphics[width=0.5\textwidth]{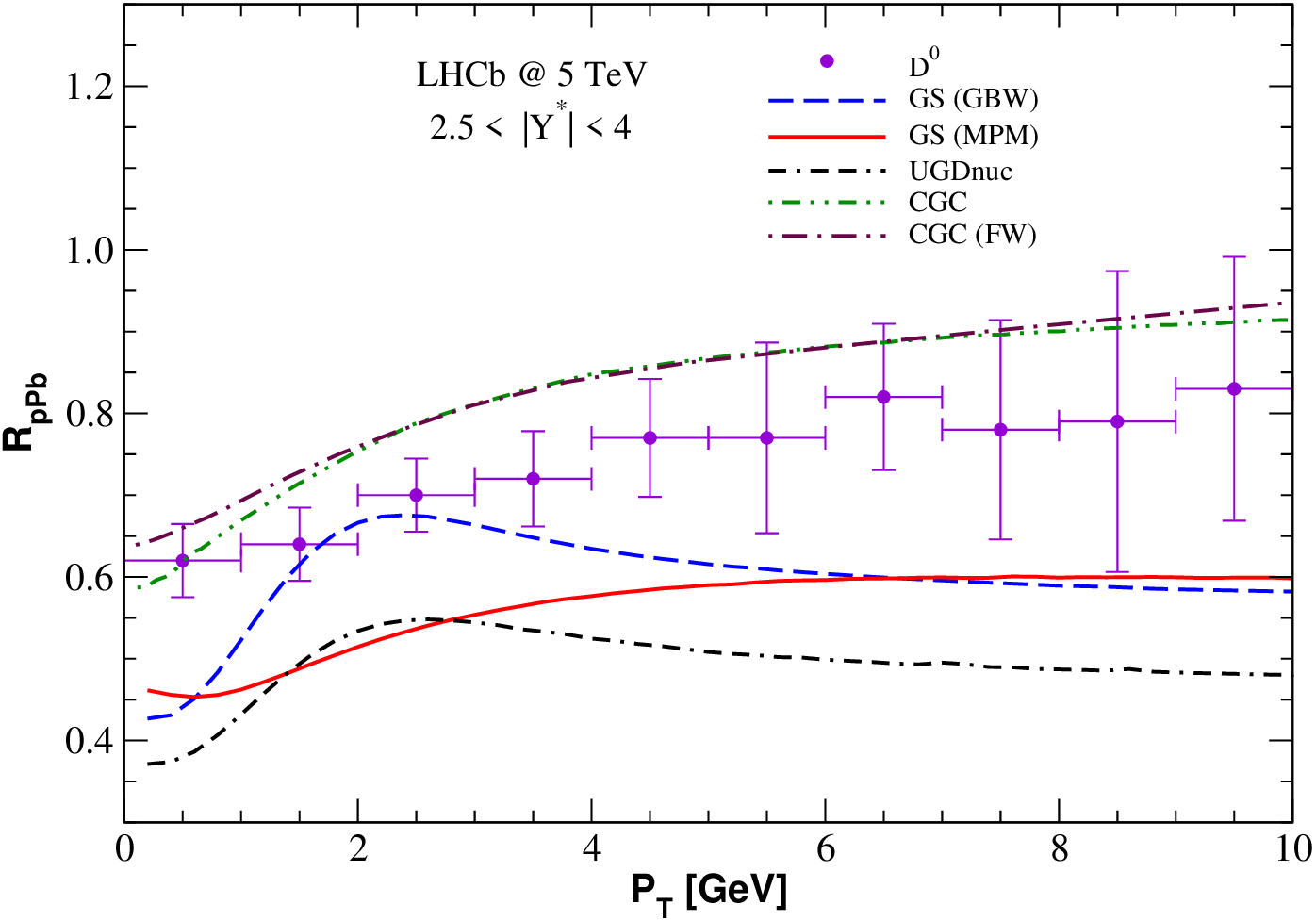}
\includegraphics[width=0.5\textwidth]{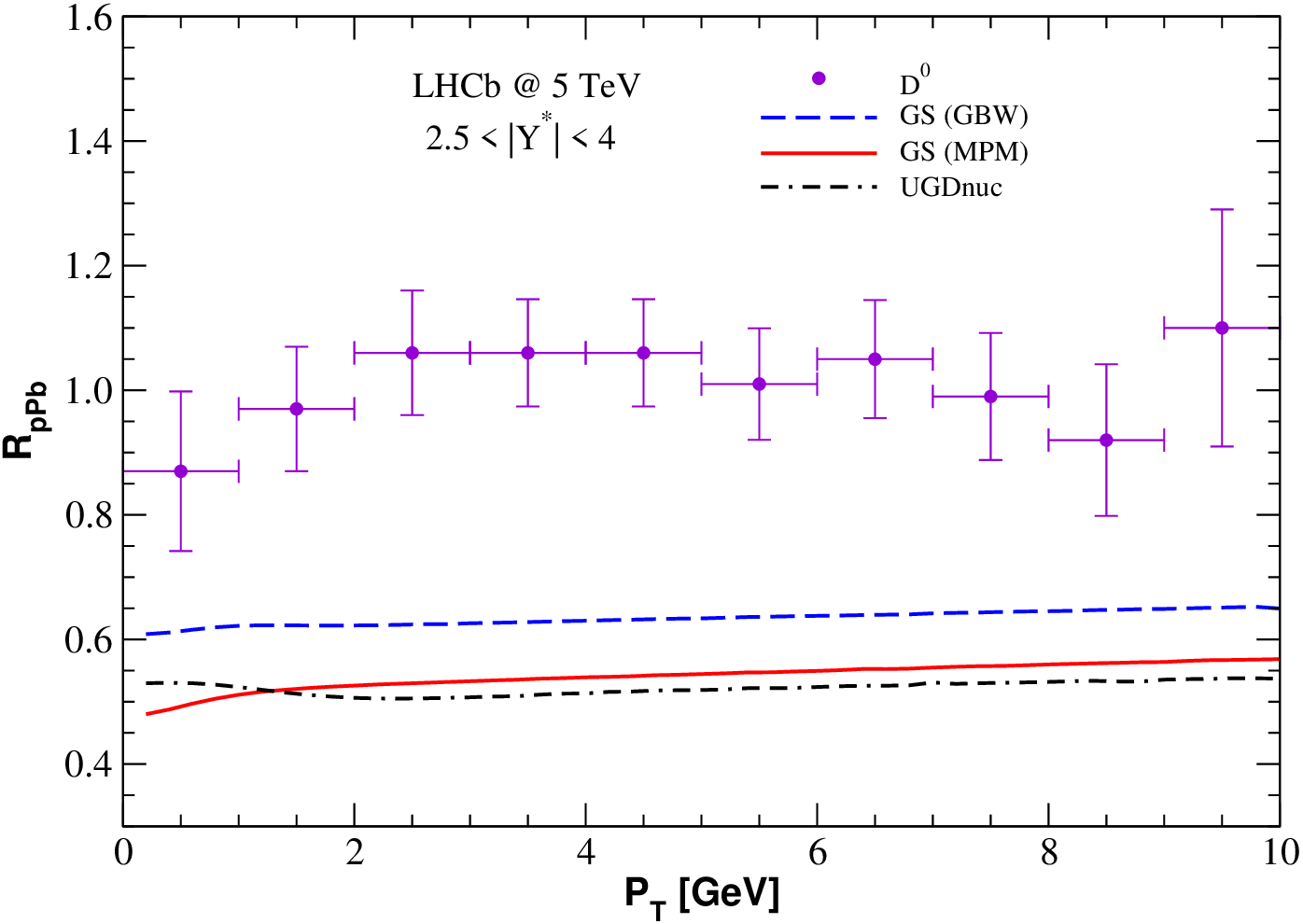}
\end{tabular}
\caption{The nuclear modification factor for $D^0$ meson production in $pPb$ collisions at $\sqrt{s} = 5$~TeV in the forward (left panel) and backward data (right panel) considering the common rapidity range, $2.5 < |Y^{*}| < 4$. In particular, the predictions given by the \emph{GS (GBW)}, \emph{GS (MPM)} and \emph{UGDnuc} models are directly compared to the CGC predictions \cite{Ducloue:2015gfa,Fujii:2017rqa} concerning the forward configuration. The experimental data are reported by the LHCb Collaboration \cite{LHCb5}.}
\label{rpPb5}
\end{figure*}

In what follows we present the results concerning the production of $D^0$, $D^+$, and $D^{*+}$ mesons assuming the rapidity interval $- 0.96 < Y^{*} < 0.04$. The theoretical predictions are shown in Fig.~\ref{pPb502} and compared against the experimental data from the ALICE Collaboration \cite{ALICE502} in $pPb$ collisions at $\sqrt{s} = 5.02$~TeV. We verify that the results obtained with the \emph{UGDnuc} and \emph{GS (GBW)} parameterizations are not in agreement with the  shape of the $P_T$ distribution when taking into account the three $D$ mesons.
Differently, the \emph{GS (MPM)} provides a better consistency, especially at large values of $P_T$. Moreover, the \emph{GS (APPROX.)} prediction is not able to reproduce the experimental measurements for the $D^0$ production. However, this scenario changes for the $D^+$ and $D^{*+}$ production considering $P_T > 8$~GeV.
Furthermore, we present the ratios obtained via the $D$-mesons production cross section in the kinematic variables at $\sqrt{s} = 5.02$~TeV and $-0.96 < Y^{*} < 0.04$. Figure~\ref{ratio} displays the following $D^+/D^0$ and $D^{*+}/D^0$ ratios as function of  $P_T$. The data were collected by the ALICE Collaboration \cite{ALICE502}.
The resulting predictions from the models are quite identical and describe the corresponding data within the experimental uncertainties. This implies that one cannot constrain the approaches by using such ratios. In other aspect, the ratios show a constant magnitude regarding the $P_T$ kinematic range, manifesting a weakly dependence on $P_T$. Consequently, those ratios do not allow us to extract particular information concerning the charm quark fragmentation functions from the $D^0$, $D^+$, and $D^{*+}$ meson decays. Moreover, the $D^0$ production rate is clear to be higher than the $D^+$ and $D^{*+}$ ones because the respective ratios are smaller than unity.

As a final investigation, we perform an analysis considering the nuclear modification factor computed by
\begin{eqnarray}
R_{pPb} = \frac{1}{A}\,\frac{d^{3}\sigma{(pA \rightarrow DX)}/dY d^2P_T}{d^{3}\sigma{(pp \rightarrow DX)}/dY d^2P_T}\,.
\label{nmf}
\end{eqnarray}
The corresponding results for $D^0$, $D^+$, and $D^{*+}$ production are shown in Fig.~\ref{rpPb} and compared to the measurements provided by the ALICE Collaboration \cite{ALICE502}.
Considering the experimental uncertainties, we notice that the results obtained with the \emph{UGDnuc} e \emph{GS (MPM)} approaches provide a better description  than the \emph{GS (GBW)} and \emph{GS (APPROX.)} ones except for the $D^+$ case, where all models are in agreement with the data. However, the \emph{GS (APPROX.)} predictions reproduce the  ratio weakly dependent on $P_T$ as discussed in previous section. Apparently, the experimental measurements for the nuclear modification factor tend to unity, $R_{pPb} \approx 1$, suggesting that the nuclear effects have no important impact for $D^0$, $D^+$, and $D^{*+}$ production at midrapidity range.

It is timely to discuss the $x_2$-values probed in the measured $P_T$ spectrum of $D$-meson production considering the kinematic range accessible at the LHCb and ALICE experiments, mainly for very forward/backward rapidity bins (see Tab.~\ref{tab}). We can verify that the values are within the validity region of the color dipole formalism, namely $x_2\leq10^{-2}$ and intermediate $P_T$. This allows us to make feasible predictions applying such approach. One exception is the configuration of very backward rapidity bin at 5~TeV in the LHCb data, although the $x_2$-value is near the validity limit.

\begin{table*}[b]
\centering
\caption{The values for the center-of-mass energy, rapidity bins, median of $P_T$, and $x_2$ for $D$-meson production in the kinematic features provided by the LHCb and ALICE measurements.}
\begin{tabular}{lccc}
\hline\hline
  $\sqrt{s_{NN}}$ (TeV) & Rapidity bin & $ P_T$ (median) & $x_2$ (median) \\
\hline
 5 (LHCb) &  $3.5 < Y^{*} < 4$         & 3.5 GeV               & $3.6\times 10^{-5}$   \\
 5 (LHCb) &  $-5 < Y^{*} < -4.5$       & 3.0 GeV                 & $1.6\times 10^{-1}$    \\
\hline 
 8.16 (LHCb) &  $3.5 < Y^{*} < 4$      & 3.5 GeV               & $2.2\times 10^{-5}$    \\
 8.16 (LHCb) & $-5 < Y^{*} < -4.5$     & 3.0 GeV                 & $9.5\times 10^{-2}$   \\
\hline
 5.02  (ALICE) & $- 0.96 < Y^{*} < 0.04$   & 18 GeV                & $1.1\times 10^{-2}$ \\
\hline\hline
\end{tabular}
\label{tab}
\end{table*}

Let us compare our calculations with other approaches in the literature. In Ref.~\cite{Eskola:2019bgf} the $D^0$ production in $pPb$ collisions has been addressed using a next-to-leading order pQCD calculation and a comparison between the input nuclear parton distribution functions (nPDFs) is done. The ratios $R_{pA}(y,P_T)$ are in good agreement with recent data using both the EPPS16 and nCTEQ15  nPDFs. Nuclear gluon shadowing at small-$x$ is predicted and it is argued that the description of a pure collinear approach is robust even at $P_T\rightarrow 0$. A study performing the reweighting of the two referred nPDFs using the LHC data for $D$, $J/\psi$, and $\Upsilon (1S)$ production in $pPb$ collisions at LHC was done in Ref.~\cite{Kusina:2020dki}. The main conclusions in work of Ref.~\cite{Kusina:2020dki} are similar to the ones achieved in Ref.~\cite{Eskola:2019bgf}. The role played by the nuclear effects driven by the fully coherent energy loss (FCEL) in cold nuclear
matter on the open heavy-flavour production has been investigated in Ref.~\cite{Arleo:2021bpv}. It has been demonstrated that the FCEL effects on $D$ and $B$ production is quite relevant and similar to those quantified in quarkonium and light hadron production. It is argued that the effect corresponds to about half of the nuclear suppression measured at the LHC at forward rapidities and low $P_T$ \cite{Arleo:2021bpv}. After inspection, our results at midrapidity are compatible with those studies.

In the context of the CGC framework, Ref.~\cite{ducloue} presents the nuclear  modification factors $R_{pA}^{J/\psi}(y,P_T)$ and $R_{pA}^{D}(y,P_T)$ at forward rapidities. The calculations make use of the Glauber model to obtain the dipole-nucleus cross section, $\sigma_{dA}(x,r,b)$. Predictions are in good agreement with the existing data at the time and the formalism had been introduced in Refs.~\cite{Ducloue:2015gfa,Ducloue:2016pqr} for quarkonium production. A distinct procedure is employed in Ref.~\cite{Fujii:2017rqa}, where transverse momentum dependent multi-point Wilson line correlators are used to describe the target nucleus and proton projectile. The corresponding UGDs are obtained from the numerical solution of the running coupling Balitsky-Kovchegov (rcBK) equation. The numerical results are consistent with data and there is a relevant dependence on the initial nuclear saturation scale, $Q_{s0,A}^2$, for the amount of nuclear shadowing appearing in $R_{pA}^D$ as a function of $P_T$ (see also Ref.~\cite{Fujii:2019yqv} for a compilation of results within the same framework for quarkonium and charged hadron production). In Fig.~\ref{rpPb5} we show our results for $R_{pPb}$ compared to those obtained by the CGC approach provided in Ref.~\cite{Ducloue:2015gfa} (denoted as \emph{CGC}) and in Ref.~\cite{Fujii:2017rqa} [referred as \emph{CGC (FW)}] taking the forward rapidity bin ($2.5 < |Y^{*}| < 4$) for $D^0$ production at $\sqrt{s} = 5$~TeV measured at the LHCb experiment. Moreover, the corresponding results considering the backward rapidity bin are shown. Predictions from \emph{GS (GBW)} (dashed curve), \emph{GS (MPM)} (solid curve) and \emph{UGDnuc} models (lower dot-dashed curve) are presented.  Our predictions contain more suppression than that predicted by the CGC ones. In particular, the \emph{GS (GBW)} result predicts less suppression compared to the results obtained by the \emph{GS (MPM)} and \emph{UGDnuc} approaches, however, they do not provide the correct normalization to describe the $P_T$-spectrum data regarding both forward and backward configuration. This can be traced back to the saturation scale: in MPM model, the proton saturation scale $Q_{s,p}^2(x=10^{-2})= 0.17$~GeV$^2$ whereas in Ref.~\cite{Ducloue:2015gfa} $Q_{s0}^2 = 0.060$~GeV$^2$ and $Q_{s0}^2 = 0.1597$~GeV$^2$ in \cite{Fujii:2017rqa}. In our case, $Q_{s,A}^2\approx 3 Q_{s,p}^2$. The backward rapidity the QCD color dipole approach can be employ as well. This is associated to the $x_2$ probed in this particular kinematic region and the validity imposed by the approach. Considering the mean values associated to the nuclear modification factor at backward rapidity: $\langle P_T\rangle~=~5$~GeV, $\langle Y\rangle~=~-3.75$ and 5~GeV of CM energy imply that $\langle x_2\rangle$ $\approx$ $4\times10^{-2}$. This is a small value for $x_2$ and is within the validity region of the color dipole formalism making such approach suitable to perform predictions. 
Still within the CGC effective theory, the absolute spectra for $D^0$, $D^+$, and $D^{*+}$ in $pp$ and $pA$ collisions have been presented in the comprehensive study of Ref.~\cite{Ma:2018bax} (see Ref.~\cite{Ma:2015sia} for similar studies addressing nuclear modification factors in quarkonium production). The focus in Ref.~\cite{Ma:2018bax} is on the so-called event engineering, which is related to the spatial and momentum structure of rare parton configurations in high-energy collisions realized by changes in the system size, multiplicity, and energy. The potential of event engineered heavy flavor measurements to reveal the dynamics of these unparalleled configurations has been demonstrated.

\begin{figure*}[!t]
\begin{tabular}{cc}
\includegraphics[width=0.5\textwidth]{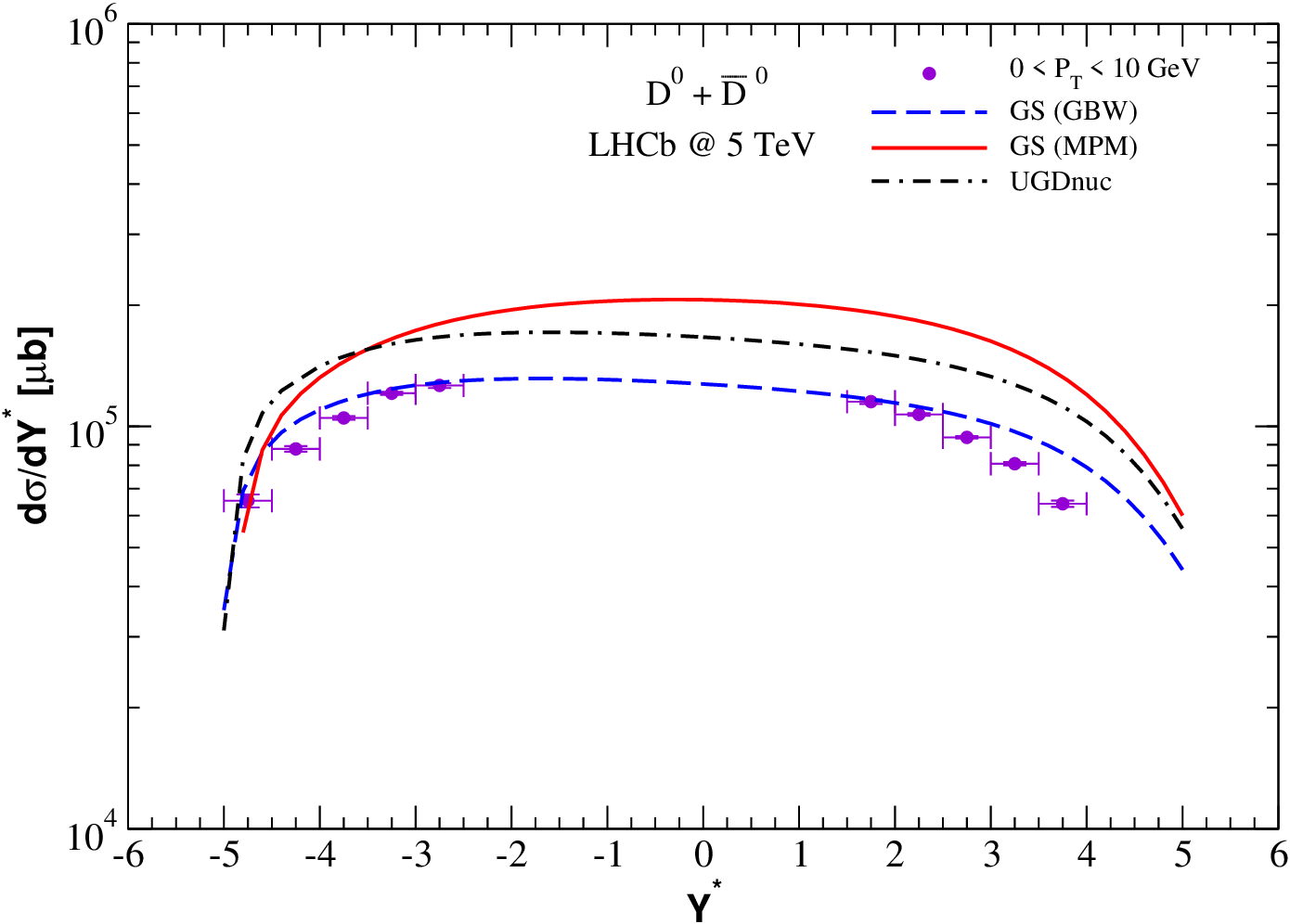}
\includegraphics[width=0.5\textwidth]{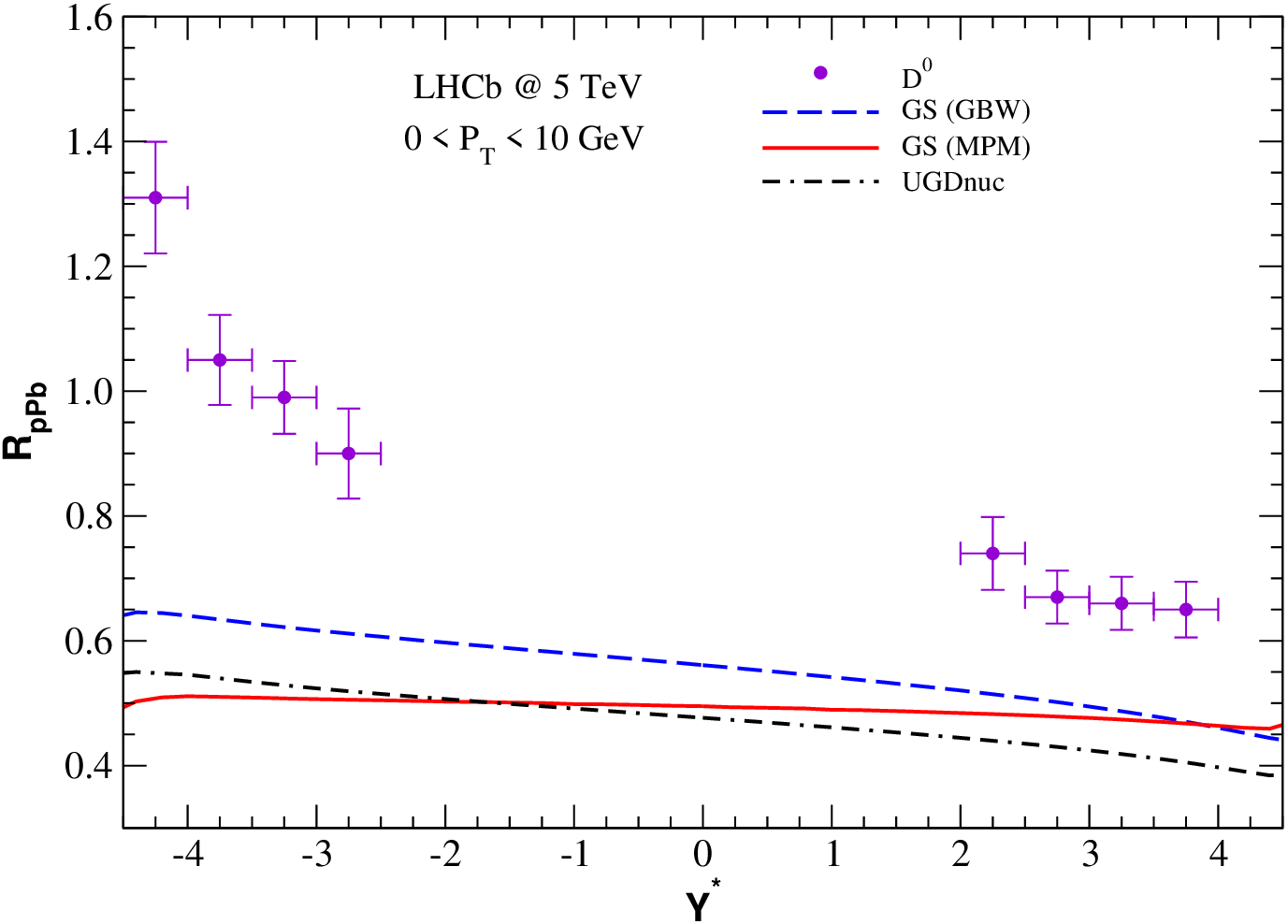}
\end{tabular}
\caption{The differential cross section (left panel) and the nuclear modification factor (right panel) for $D^0$ meson production in $pPb$ collisions at $\sqrt{s} = 5$~TeV as a function of $Y^{*}$ in the forward and backward configuration. A comparison between the predictions with the \emph{GS (GBW)}, \emph{GS (MPM)} and \emph{UGDnuc} models and the data reported by the LHCb Collaboration \cite{LHCb5} is performed.}
\label{xs_rpPb5}
\end{figure*}

Finally, we present the results concerning the rapidity dependence of the cross section and the nuclear modification factor of $D$-meson integrated over $P_T$, $0~<~P_T~<~10$~GeV. The predictions for $D^0$ production at 5~TeV that are compared to the experimental points provided by the LHCb Collaboration are shown in Fig.~\ref{xs_rpPb5}. Apparently the models delivered a better description of the cross section taking the negative rapidity bin. In particular, the GBW model provides a better agreement concerning the data description at positive rapidity bin. In general, considering the rapidity distribution, the results overshoot the data. This is expected as we can see the results taking the $P_T$ spectrum, $0<P_T<2$~GeV, in Fig.~\ref{pPb5}, where the corresponding theoretical predictions overestimate the experimental data. On the other hand, the results for the nuclear modification factor show that $R_{pPb}<1$ with a decreasing towards to positive rapidity, i.e., forward configuration. This can be associated to the higher rate of $D^0$ production in $pp$ collisions as we can observe from Fig.~\ref{pp5_13} and it directly affects the values of $R_{pPb}$, see Eq.~\ref{nmf}. The percentile value of the rate of $D^0$ production that exceeds the experimental measurements in $pp$ collisions is approximately 30$\%$.

\begin{figure*}[!t]
\begin{tabular}{cc}
\includegraphics[width=0.5\textwidth]{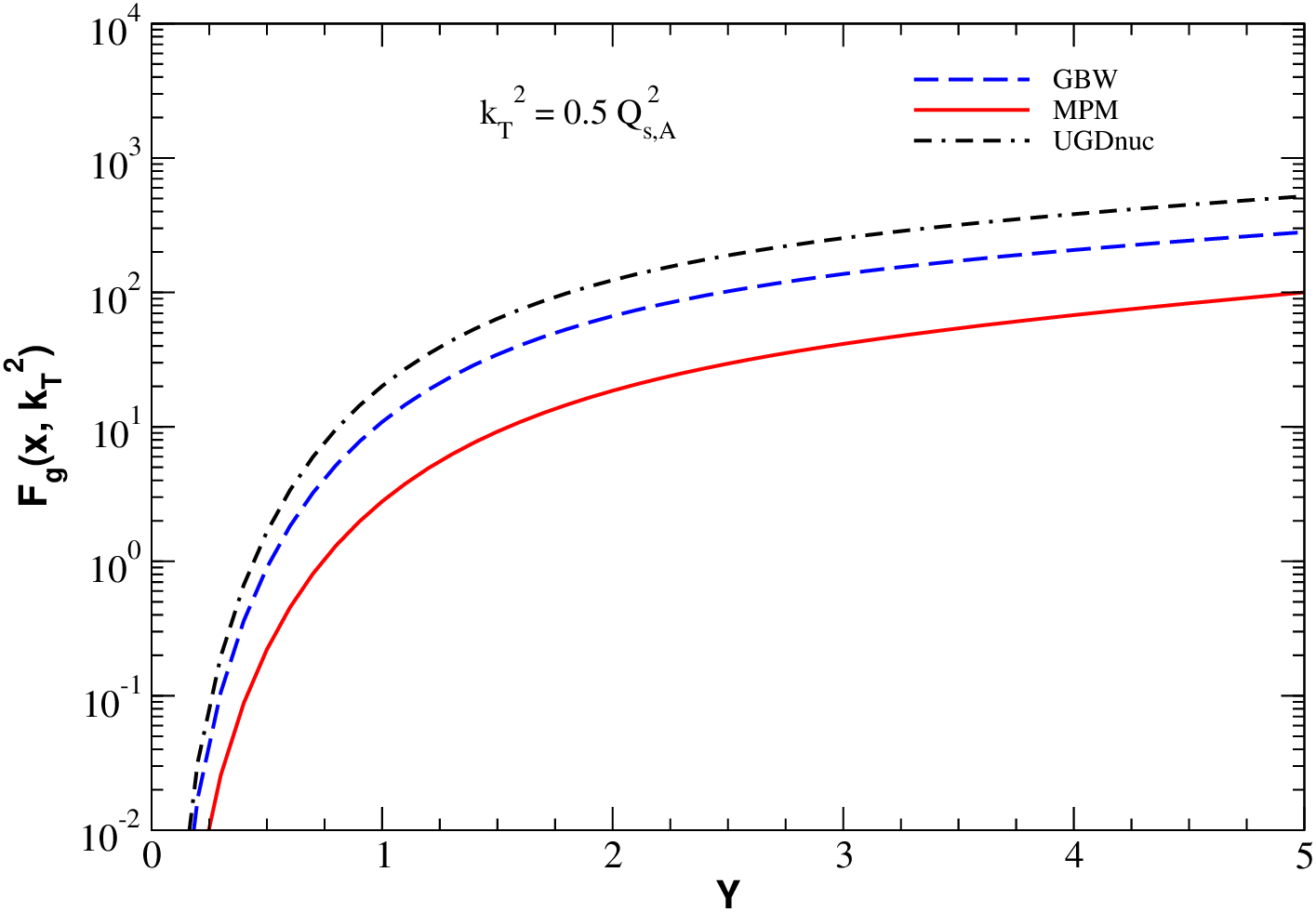} \includegraphics[width=0.5\textwidth]{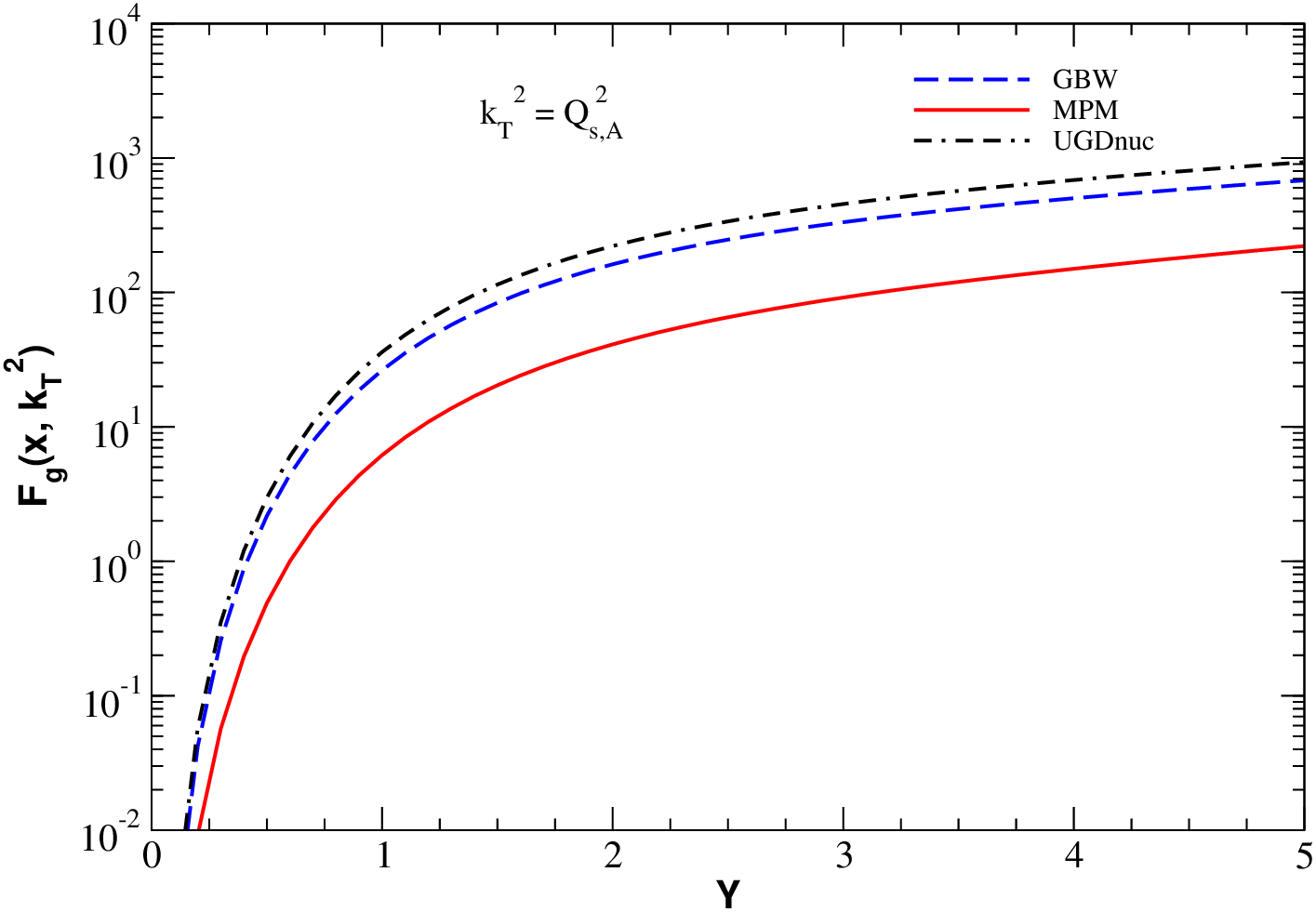}\\
\includegraphics[width=0.5\textwidth]{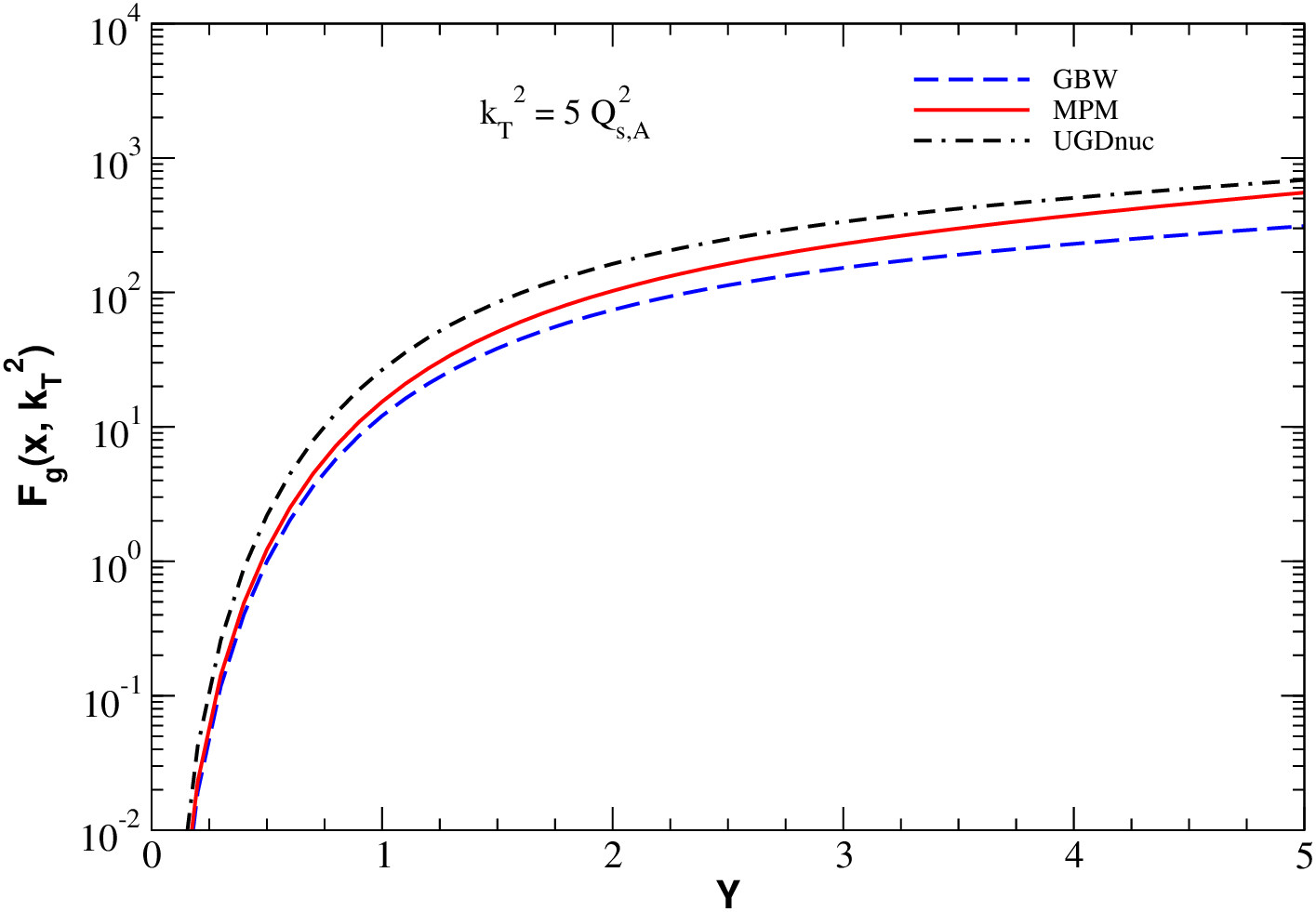} 
\end{tabular}
\caption{The rapidity dependence of the UGD parameterizations evaluated at the transverse momentum values given by $k^{2}_T = 0.5\, Q^{2}_{s,A}$, $k^{2}_T = Q^{2}_{s,A}$,  and $k^{2}_T = 5\, Q^{2}_{s,A}$.}
\label{ugdy}
\end{figure*}

For the sake of completeness, we show the rapidity dependence of the UGD parametrizations. Indeed, the $x_2$ dependence of the UGD allows us to investigate the rapidity dependence of the $D$-meson production. The results considering three values of the transverse momentum in terms of the nuclear saturation scale, $k^{2}_T = 0.5\, Q^{2}_{s,A}$, $k^{2}_T = Q^{2}_{s,A}$, and $k^{2}_T = 5\, Q^{2}_{s,A}$, are presented in Fig.~\ref{ugdy}. The predictions present a slightly difference starting from the central rapidity ($Y=0$) and this difference becomes more pronounced as the rapidity evolution increases. In particular, the MPM result gives a larger deviation in comparison to the others parametrizations, except from $k^{2}_T = 5\, Q^{2}_{s,A}$, where the deviation between the UGD predictions turn less apparent. Consequently, this difference observed with GBW, MPM, and UGDnuc parametrizations is explicitly converted into the results predict for the cross section and the nuclear modification factor of the $D$-meson production.

\section{Summary} 
\label{conc}

In this work we perform an analysis concerning the $D^0$, $D^+$, and $D^{*+}$ production in $pPb$ collisions in the high-energy limit considering the kinematic region achieved at the CERN-LHC. The predictions are computed by applying the color dipole approach in transverse momentum representation with the GBW and MPM unintegrated gluon distribution including the geometric scaling property. Results for the nuclear gluon distribution based on Glauber-Gribov theory and  an approximated expression valid at $P_T>Q_{s,A}$, Eq.~(\ref{approximation}), are presented as well. We have found that the predictions obtained with the MPM parameterization in conjunction with the geometric scaling are very consistent with  measurements reported by the ALICE and LHCb Collaborations over  a wide $P_T$ range in forward and backward rapidity bins. Nonetheless, we can not distinguish between the approaches by means of the ratios for $D^0$, $D^+$, and $D^{*+}$ production. The magnitude of nuclear effects associated to the production of $D$-mesons in $pPb$ collisions seems to be small at not so small-$P_T$ at midrapidities, since the nuclear modification factor measured is consistent with unity given the experimental uncertainties.

The color dipole in transverse momentum representation offers a suitable framework to evaluate the $D$-meson production and effective in forthcoming investigations of $D$-meson measurements in heavy-ion programmes. Moreover, more data from future experimental measurements on nuclear modification factor at forward and backward rapidities are needed to further constrain the different approaches. Thus, one will be able to refine the associated phenomenology as well as the assumptions encoded in the UGDs. We restrict our investigations to analytical expressions for the UGD in protons/nuclei which parametrize the parton saturation effects. Future studies considering the numeric solutions of the nonlinear evolution equations, as the running coupling BK equation, would be valuable. 

\section*{Acknowledgements}

We are grateful to Giulia Manca (Università degli studi di Cagliari) for  valuable feedback on the LHCb data for $D$ meson production in $pA$ collisions. This work was partially financed by the Brazilian funding agencies CAPES, CNPq, and FAPERGS. This study was financed in part by the Coordena\c{c}\~ao de Aperfei\c{c}oamento de Pessoal de N\'{\i}vel Superior - Brasil (CAPES) - Finance Code 001. GGS
acknowledges funding from the Brazilian agency Conselho Nacional
de Desenvolvimento Científico e Tecnológico (CNPq) with grant
311851/2020-7.

\end{document}